\documentclass[aps,pre,twocolumn,notitlepage,showpacs,10pt]{revtex4-1}

\usepackage{bm}
\usepackage{graphicx}
\usepackage{latexsym}
\usepackage{amsmath}
\usepackage[colorlinks=true,citecolor=blue,linkcolor=blue]{hyperref}
\usepackage{bbding}
\usepackage[caption=false]{subfig} 
\usepackage{float}
\usepackage{subfloat}
\usepackage[section]{placeins}
\usepackage{csquotes}
\usepackage{relsize}
\usepackage{enumitem}
\DeclareGraphicsExtensions{.pdf,.png,.jpg}

\LetLtxMacro{\OldSqrt}{\sqrt}
\newcommand{\ClosedSqrt}[1][\hphantom{3}]{\def\DHLindex{#1}\mathpalette\DHLhksqrt}
\makeatletter
    \newcommand*\bold@name{bold}
    \def\DHLhksqrt#1#2{%
        \setbox0=\hbox{$#1\OldSqrt{#2\,}$}\dimen0=\ht0\relax%
        \advance\dimen0-0.2\ht0\relax
        \setbox2=\hbox{\vrule height\ht0 depth -\dimen0}%
        {\hbox{$#1\expandafter\OldSqrt\expandafter[\DHLindex]{#2\,}$}
        \lower\ifx\math@version\bold@name0.6pt\else0.4pt\fi\box2}
    }
    \renewcommand*{\sqrt}[2][\ ]{\ClosedSqrt[\leftroot{-2}\uproot{1}#1]{#2}\kern0.1em} 
\makeatother

\renewcommand\vec{\mathbf}

\begin{document}

\title{Strongly-Mismatched Regime of Nonlinear Laser-Plasma Acceleration: \\
Optimization of Laser to Energetic Particle efficiency}

\author{Aakash A. Sahai}
\affiliation{College of Engineering and Applied Science, University of Colorado, Denver, CO 80231}
\email[corresponding author: ~]{aakash.sahai@gmail.com}

\begin{abstract}
A strongly mismatched regime of self-guided nonlinear laser-plasma acceleration in the bubble regime is modeled for optimization of Laser to Particle energy efficiency with application to recently proposed laser positron accelerator. The strong mismatch, in contrast with the matched condition, arises from the incident laser spot-size being much larger than that needed for equilibration of the laser ponderomotive and electron-ion charge-separation force in the plasma bubble. This is shown to be favorable for optimization of large self-injected electron charge and ultra-low transverse emittance. The prominent signatures of the mismatched regime, strong optical-shock excitation and bubble elongation, are validated using multi-dimensional Particle-In-Cell simulations. This work thus uncovers a generalized regime that is shown to have been favored by many laser-plasma acceleration experiments and opens a novel pathway for a wide-range of future applications. 
\end{abstract}

\maketitle

\section{Introduction}\label{sec:introduction}
Laser-plasma accelerators (LPA) \cite{Tajima-Dawson} using self-guided laser-driven non-linear electron density waves in the \enquote{bubble} regime \cite{Pukhov-bubble-2002} that enabled the ground-breaking centimeter-scale acceleration \cite{Mangles-2004,Faure-2004} of electron \enquote{beams} have now inspired a worldwide effort on LPAs. These efforts on self-guided LPA \cite{UT-Austin, Nebraska, GIST, Strathclyde} have continued to show enhancement in the electron beam properties.

The theoretical model of these bubble LPAs is based on a \enquote{matched} regime \cite{Pukhov-bubble-2002}. Theoretically, maximum energy gain is considered to be only possible if the incident laser radial spot-size is matched to the \enquote{bubble} size that equilibrates the electron-ion charge-separation and the laser ponderomotive force. This initially matched laser spot-size condition is held to be exclusively optimal for properties of the acceleration structure and electron beam \cite{Lu_PRSTAB_2007}.

However, a wide-range of well-known groundbreaking experimental results \cite{Mangles-2004, Faure-2004, UT-Austin, Nebraska, GIST, Strathclyde} that have established these LPAs have favored a strongly mismatched regime. In the strongly mismatched regime, modeled here for the first-time, the laser focal spot-size is significantly larger than the matched condition. 

This work seeks to optimize the Laser to Energetic particle ($\rm\geq 1.02 MeV$) conversion efficiency to enable applications such as the feeder-stage of recently proposed laser positron accelerator \cite{positron-LPA} as opposed to the energy spread of the particle beam. Past experimental work has often used this regime because it has also been found to be more effective for peak energy gain in comparison with the matched regime. Despite the higher electron beam energy and other qualities which experimentally establish the profound importance of the mismatched regime, no earlier work has investigated its underlying physical mechanisms.

Physical processes underlying the mismatched regime are here shown to significantly differ from the matched regime. Two prominent signature processes of this regime - {\it strong optical-shock excitation} and {\it bubble elongation} are elucidated here. The process of laser slicing uncovered here is significantly different from the well-known effect of laser etching in plasma. Thus, while this work reveals novel laser-plasma dynamics of mismatch it also opens up an alternative to the matched regime. 

Apart from merely the tendency of experiments to favor the mismatched regime, several important factors motivate its study. Firstly, a larger vacuum focal spot-size at a given laser power is known to produce higher \enquote{mode-quality} in the far-field (low beam-propagation factor or $\mathcal{M}^2$-number or $\rm{TEM_{00}}$-times diffraction-limited number). This is because a larger spot is less affected by various aberrations \cite{Siegman-1997}. High mode-quality is not equivalent to the maximization of intensity percentage within the focal-spot, as characterized by the Strehl ratio \cite{Strehl-OL-1998}. This is because of the well-defined \enquote{no-$\rm TEM_{00}$ gaussian} problem \cite{Siegman-1997}. Secondly, self-injection mechanisms which rely on symmetry breaking processes, promise beams with distinctive phase-space properties and higher net charge due to the inherent mismatch \cite{Max-1974, Sprangle-1987, Sun-1987, Hafizi-2000, KV-1959, Kalmykov-2009}.

Ideal laser focal-spot mode characteristics are assumed in the bubble regime theory of self-guided LPAs \cite{Pukhov-bubble-2002}. A laser with predominantly $\rm TEM_{00}$-mode is self-guided in a homogeneous plasma with electron density, $n_0$ over multiple Rayleigh lengths, $Z_R\equiv\pi w_0^2/\lambda_0=\pi W_0^2/(\mathcal{M}^2\lambda_0)$ ($\lambda_0$ is the central wavelength of the laser and $w_0$ is the spot-size of $\rm TEM_{00}$ mode and $W_0$ is the measured spot-size) if it drives electron cavitation \cite{Sprangle-1987,Sun-1987}. To self-guide the $\rm TEM_{00}$-mode, the laser power $P$ has to exceed a critical power $P_c = 17.4\times 10^9 ~ (\omega_0/\omega_{pe})^2 ~ \rm W$ ($\omega_0$ is laser frequency and $\omega_{pe}=\sqrt{4\pi n_0e^2/m_e}$ is the plasma frequency). When $P\geq P_c$, the plasma refractive index profile is shaped by relativistic-quiver \cite{Sprangle-1987,Sun-1987} and ponderomotive-channeling \cite{Hafizi-2000} which counter the diffraction of the focused laser. 

Focused laser modal composition characteristics are constrained as the bubble plasma wave excitation demands the peak normalized laser vector potential, $a_0\gg 1$ \cite{Pukhov-bubble-2002} (where $a_0=\rm{max}[e\vec{A}/m_ec^2]=8.55\times10^{-10}\lambda_0[\mu m]\sqrt{I_0[\rm{W/cm^2}]}$, $\vec{A}$ is the laser vector potential, $I_0$ is the peak intensity and $\lambda_0$ the wavelength) which typically also satisfies $P\geq P_c$. Importantly, the bubble spatial profile is dictated by the laser focal mode characteristics which under the required tight focussing suffer from optics induced aberrations and distortions \cite{Siegman-1997}.

In the bubble regime, the electron-ion charge separation equilibrates with the laser ponderomotive force at the matched spot-size \cite{Lu_PRSTAB_2007}, 
\begin{equation}
\begin{aligned}
w_{\rm 0-m}  = 2 \sqrt{a_0} ~ \frac{c}{\omega_{pe}} = R_{\rm{bubble}}
\end{aligned}
\label{eq:radial-matched-condition}
\end{equation}
\noindent and thus the matched initial conditions enforce a vacuum laser spot-size with $w_{\rm 0-m} = R_{\rm bubble}$. 

The electron energy gain in this matched regime based upon 3D PIC simulations scales as \cite{Lu_PRSTAB_2007}:
\begin{equation}
\begin{aligned}
\Delta \mathcal{E} ~ [m_ec^2] \simeq \frac{2}{3} ~ a_0 ~ \left(\frac{n_c}{n_0}\right)
\end{aligned}
\label{eq:radial-matched-condition-energy}
\end{equation}
\noindent In eq.\ref{eq:radial-matched-condition-energy}, $a_0$ is the vacuum vector potential that is incident on the plasma. The value of $a_0$ in plasma is known to significantly vary over the acceleration length due to several laser-plasma interactions effects. These include localized variation of the wavelength profile, group velocity profile and pump depletion of the laser pulse in addition to radial self-focussing. Thus, this equation best models a scenario where $a_0$ is stable over the acceleration length, as is argued to be the case for matched spot-size.

However, contrary to the matched regime theory exemplary experimental evidence \cite{Mangles-2004, Faure-2004, UT-Austin, Nebraska, GIST, Strathclyde} has shown optimality of mismatched regime. In these experiments, density scans with a fixed laser spot-size have shown that the beam energy gain and other properties optimize where incident spot-size ($W_0$) is much larger than the matched spot-size ($w_{\rm 0-m}$). A mismatch factor, $\Gamma$ is defined as:
\begin{equation}
\begin{aligned}
\Gamma = W_0/w_{\rm 0-m} \gg 1.
\end{aligned}
\label{eq:mismatch-factor}
\end{equation}
\noindent A large mismatch inflates the difference between eq.\ref{eq:radial-matched-condition-energy} and experiments (sec.\ref{sec:adjusted-a0-model}). Thus the analysis here is based on GeV-scale energy gain data \cite{Poder-2016,Kneip-2009} that have shown maximum beam energy gain in a mismatched regime.

The first bubble wave \cite{Pukhov-bubble-2002} based self-guided experiments used the mismatched regime to demonstrate laser-plasma acceleration of quasi-mono-energetic electron beams \cite{Mangles-2004,Faure-2004}. In \cite{Mangles-2004}, the incident laser intensity was $\rm 2.5\times 10^{18} Wcm^{-2}$ ($a_0$=1.1) at $\rm n_0 = 2\times 10^{19} cm^{-3}$, the matched spot-size is $w_{\rm 0-m}\rm\simeq 3\mu m$ whereas the launched spot-size was $W_0 \simeq \rm 12\mu m$ (FWHM $\simeq \rm 20\mu m$), a mismatch of $\Gamma \simeq 4$. The predicted energy from eq.\ref{eq:radial-matched-condition-energy} is 40 MeV but experiments obtained a spectral peak at 70 MeV. Similarly in \cite{Faure-2004}, the intensity was $\rm 3.2\times 10^{18}Wcm^{-2}$ ($a_0$=1.3) at $\rm n_0 = 6\times 10^{18} cm^{-3}$, the matched spot-size is $w_{\rm 0-m}\rm\simeq 5\mu m$ whereas the launched spot-size of $W_0=\rm 12.5\mu m$ (FWHM $\rm\simeq 21\mu m$), a mismatch of $\Gamma \simeq 2.5$. The expected beam energy is 155 MeV but the spectral peak was at 175 MeV. 

Moreover, the mismatched regime attains higher relevance as it has continued to be applicable to many other ground-breaking experiments that have driven LPAs forward. Some prominent examples are, Austin-2 GeV data \cite{UT-Austin}: $W_0=275\mu m$, $a_0=0.6$, $n_0=5\times 10^{17} ~ \rm{cm^{-3}}$, $w_{\rm 0-m}\simeq 12\mu m$, $\Gamma = 22$; Nebraska-0.3 GeV data \cite{Nebraska}: $ W_0=17 \rm{\mu m}$, $a_0=2.2$, $n_0=2.5\times 10^{18} ~ \rm{cm^{-3}}$, $w_{\rm 0-m}\simeq 10 \rm{\mu m}$, $\Gamma \sim 2$;  Gwangju-2 GeV data \cite{GIST}: $W_0=25.5 \rm{\mu m}$, $a_0=5$, $n_0=1.4\times 10^{18} ~ \rm{cm^{-3}}$, $w_{\rm 0-m}\simeq 20 \rm{\mu m}$, $\Gamma \sim 1.3$ and Strathclyde-125MeV data \cite{Strathclyde}: $W_0=20 \rm{\mu m}$, $a_0=1.5$, $n_0=1\times 10^{19} ~ \rm{cm^{-3}}$, $w_{\rm 0-m}\simeq 5 \rm{\mu m}$, $\Gamma = 4$. 

Further experimental evidence in favor of a larger than matched-size focal spot also comes from observation of the reduction of experimental artifacts which affect the plasma-wave quality. A non-uniform laser focal-spot (a large $\mathcal{M}^2$-number) is known to affect the transverse characteristics of the plasma wave \cite{Faure-2015} and leads to non-optimal acceleration and focusing field profiles in the plasma. This quality degradation is in addition to a faster laser energy loss due to the tendency of higher-order modes in a focused laser spot to diffract faster. The plasma wave quality has been indirectly inferred using laser focal profile, such as the presence of multiple hot-spots, at the plasma entrance \cite{UT-Austin} and exit \cite{Kneip-2009}.

In this work, the essential dynamics due to the mismatch is analytically modeled in section \ref{sec:nonlinear-envelope-equation} with a nonlinear envelope equation of a self-guided laser in a bubble plasma wave derived below. This equation shows that the oscillations of the spot-size increasingly become asymmetric in response to an increasing degree of mismatch, with shorter (and tighter) \enquote{squeeze} phases and longer \enquote{relaxation} phases of the laser spot-size. This behavior is similar in characteristics to a \enquote{cnoidal} wave, whose form is given by the Jacobi elliptic function, \enquote{cn} \cite{Jacobi-elliptic-functions} (also solution of Korteweg-de-Vries equation \cite{KdV-eq}). Such behavior is not modeled in earlier analyses of the evolution of a self-guided laser envelope \cite{Max-1974, Sprangle-1987, Sun-1987, Hafizi-2000, KV-1959}. 

The peak intensities reached in the squeeze phase are many times higher than that in the matched regime with the same laser energy. In section \ref{sec:adjusted-a0-model} a heuristically derived adjusted-$a_0$ model shows that the highest electron energies in the strongly mismatched regime exceed the predictions of eq.\ref{eq:radial-matched-condition-energy}. Further, the rate of change of intensity in the shortened squeeze phase is unprecedented. Consequently, the laser-plasma interaction processes in the rapid spot-size squeeze events dictate the physical mechanisms that underlie this regime.

By an analysis of the physical mechanism that underlie this regime using multi-dimensional PIC simulations in section \ref{sec:pic-simulations} it is here shown that the laser spot-size mismatch leads to processes yet unexplored.

The {\it general applicability} of the strongly mismatched regime is here proved using 3D and 2.5D Particle-In-Cell (PIC) simulations. The {\it validation of methodology} of computational modeling is proved using the equivalence of 3D PIC simulations with boosted vacuum-$2\times a_0$ 2.5D PIC simulations for experimental data in \cite{Mangles-2004}.

Using 3D and 2.5D PIC simulations in section \ref{sub-sec:analysis-PIC-simulation} the two signature processes of the mismatched regime are validated. A rapid rise in the intensity in the squeeze phase is shown in section \ref{sub-sec:strong-optical-shock} to lead to rising ponderomotive force which drives a sharply rising density perturbation ahead of the peak of the laser. The interplay of density build-up and relativistic reduction of plasma frequency in the region of increasing intensity modifies a part of the laser which overlaps with the electron density build-up, in the front of the bubble. A rapid shift in the group velocity of a part of the laser pulse leads to the slicing of the laser longitudinal envelope. This slicing sets the pulse into a state of a strong \enquote{optical shock} \cite{optical-shock} as shown in section \ref{sub-sec:strong-optical-shock}. Secondly, the slicing breaks the oscillatory envelope mode due to the coupling of the transverse to longitudinal envelope dynamics. 

This pulse slicing effect significantly differs from the slow pulse etching \cite{etching-model-1996} which sharpens the laser front. The process of etching in the matched regime is due to the red-shift of the pulse front. Slicing on the other hand essentially detaches the head of the pulse from the rest of its body and excites a strong optical shock.

The effect of a strong optical shock on acceleration mechanism is shown to be optimal only over a narrow range of densities (Fig.\ref{Fig5:laser-wavelength-a0-evolution-f40} in sec.\ref{sec:pic-simulations}). At the lower end of this range, the radial squeeze is slow and a weak optical shock is reached close to the laser energy depletion length. At the higher end, the radial squeeze is too short to allow significant interaction between the injected beam and the short-lived peak field. The envelope oscillation wavelength is also short and results in closely spaced successive laser slicing events. Both these effects inhibit acceleration to the highest energies at higher densities.

In response to the sliced pulse which forms a strong optical shock, the bubble rapidly elongates and drives a novel self-injection mechanism. The properties of the beam injected during this elongation are modeled using PIC-based analysis in section \ref{sub-sec:beam-properties}.

\section{Nonlinear envelope equation - \\ Asymmetric evolution in Bubble}
\label{sec:nonlinear-envelope-equation}

An analytical model of the evolution of the laser envelope size under laser-excited electron response using individual ray equations of geometric optics in a homogeneous plasma was first modeled in \cite{Sprangle-1987}. An envelope equation of the variation of radial envelope size $R_s(z)$ with $z=ct$, was derived in this work (size was defined as the root-mean-square of the radial location of individual rays). It was shown to be of the form of the evolution equation of a coasting particle beam \cite{KV-1959}. 

This equation which assumes radial symmetry was then simplified using the source dependent expansion with Laguerre-Gaussian eigen-functions, under the assumption, $a_{\rm m=0}\gg a_{\rm m>0}$ (where m is the mode number and represents the order of the Laguerre polynomial and $m=0$ corresponds to $\rm TEM_{00}$-mode). The equation thus obtained was then further simplified to determine the self-focusing critical power for $\rm TEM_{00}$-mode under the asymptotic approximation of a large value of $R_s(z)/(a_{\rm inc}R_{\rm inc})$, where $R_{\rm inc}$ is the incident root-mean-square envelope size and $a_{\rm inc}$ is the peak incident normalized vector potential. 

In \cite{Sun-1987}, a quasi-static approximation of a driven wave equation in vector potential was used to show that the critical laser power corresponds to the condition of complete expulsion of all electrons from within the laser volume, referred to as electron cavitation.

Envelope behavior in self-guided regime is modeled using eq.\ref{eq:Sprangle-envelope-equation}. In this equation, vacuum diffraction is balanced by laser driven refractive index profile \cite{Sun-1987}. However, this predicts that the envelope undergoes a catastrophic radial collapse for $P>P_c$ and was thus not useful above the threshold.
\begin{equation}
\begin{aligned}
& \frac{a_{\rm inc}R_{\rm inc}}{R_s(z)} \rightarrow 0 \\
& \frac{d^2 R_s(z)}{dz^2} = \frac{R_{\rm inc}^2}{a_{\rm inc}^2Z_R^2} ~ \left[ \frac{1}{R_s(z)}  \left(\frac{a_{\rm inc}R_{\rm inc}}{R_s(z)}\right)^2 \left(1 - \frac{P}{P_c} \right) \right]
\end{aligned}
\label{eq:Sprangle-envelope-equation}
\end{equation}

The above result was obtained using an insightful effective potential approach that modeled the behavior of the envelope by using a simplified \enquote{single particle} model. The location of the particle is set to $R_s$ and the right-hand-side of the envelope equation eq.\ref{eq:Sprangle-envelope-equation} is written as a normalized effective potential V($R_s,P/P_c$), to model the oscillations of this \enquote{particle} as in eq.\ref{eq:potential-envelope-equation} \cite{Sprangle-1987}\cite{Hafizi-2000}.
\begin{equation}
\begin{aligned}
\frac{d^2 R_s(z)}{dz^2} & = - \frac{R_{\rm inc}^2}{a_{\rm inc}^2Z_R^2} ~ \left( \frac{\partial V_{\rm eff}}{\partial R_s} \right) \\
V_{\rm eff}(R_s,P/P_c) & =  \frac{1}{2} ~ \left(\frac{a_{\rm inc}R_{\rm inc}}{R_s}\right)^2 \left[ 1 - \frac{P}{P_c} \right]
\end{aligned}
\label{eq:potential-envelope-equation}
\end{equation}

A different equation was derived in the opposite asymptotic limit of $P/P_c>1$ and small value of $R_s(z)/(a_{\rm inc}R_{\rm inc})$ of the form in eq.\ref{eq:Sprangle-envelope-small-x}. In this limit, the laser envelope begins to diffract when the spot-size satisfies the condition that $R_s < a_{\rm inc} R_{\rm inc} / (16 ~ P/P_c )^{1/3}$ the rate of change of envelope becomes too high for the relativistic effects to continually focus it down. Under this condition the envelope either oscillates or diffracts away depending upon its initial rate of change. However, eq.\ref{eq:Sprangle-envelope-small-x} is not applicable to the bubble regime.
\begin{equation}
\begin{aligned}
\frac{R_s(z)}{a_{\rm inc}R_{\rm inc}} & \rightarrow 0 \\
\frac{d^2 R_s(z)}{dz^2} & = \frac{R_{\rm inc}^2}{a_{\rm inc}^2Z_R^2} ~ \left( \frac{1}{R_s(z)}  \left(\frac{a_{\rm inc}R_{\rm inc}}{R_s}\right)^2 \cdots \right. \\
& \left. \qquad \qquad \qquad \left[1 - 16\frac{P}{P_c}\left(\frac{R_s}{a_{\rm inc}R_{\rm inc}}\right)^3 \right] \right)
\end{aligned}
\label{eq:Sprangle-envelope-small-x}
\end{equation}

The effective potential on a particle approach assumes that the laser spot-size is equal to the bubble radius, $R_s\simeq R_B$. PIC simulations validate this assumption nearly over the entire length of the evolution of the laser to within a small factor of the order of unity. The sheath that surrounds the bubble cavity is effectively the particle in this nonlinear \enquote{single particle} model. In the bubble regime, the condition that $P \gg P_c$ and $a_0 \gg 1$, is generally satisfied and therefore the effective potential is not dominated by the interplay between the relativistic focusing and vacuum diffraction. In consideration of this catastrophic collapse predicted for $P > P_c$ in eq.\ref{eq:potential-envelope-equation}, these terms are retained only when $P \leq P_c$.

In the nonlinear bubble regime, the effective potential is dictated by interplay between the ponderomotive potential and the ion electrostatic potential. Thus, for $P \geq P_c$, the effective potential is the sum of the ponderomotive potential of the laser pulse in eq.\ref{eq:laser-ponderomotive-potential}, 
\begin{equation}
\begin{aligned}
\phi_p(P,R_s) & = \Phi_p(P,R_s)/(m_ec^2) \\
& = \gamma_{\perp}(P,R_s) - 1 = \sqrt{1+a^2(P,R_s)} - 1
\end{aligned}
\label{eq:laser-ponderomotive-potential}
\end{equation}
\noindent where $a$ is the normalized vector potential and the ion cavity potential in eq.\ref{eq:ion-cavity-potential} on the \enquote{particle} which represents the electron sheath that surrounds the laser pulse, 
\begin{equation}
\begin{aligned}
\phi_{cav}(R_s) & = \Phi_{\rm cav}(R_s)/(m_ec^2) 
& = k_{pe}^2R_s^2/4 
\end{aligned}
\label{eq:ion-cavity-potential}
\end{equation}
\noindent where, $k_{pe}=\omega_{pe}c^{-1}$). Thus the total effective potential is as in eq.\ref{eq:nonlinear-eff-potential} and the corresponding non-linear envelope equation in eq.\ref{eq:nonlinear-envelope-eq}. Here, $\mathcal{H}$ is the Heaviside step function.
\begin{equation}
\begin{aligned}
V_{\rm eff}&\left(R_s,\frac{P}{P_c}\right)  & \\
& = \frac{1}{2} ~ \left(\frac{a_{\rm inc}R_{\rm inc}}{R_s}\right)^2 \left[ 1 - \frac{P}{P_c} \right] ~ \mathcal{H}(P_c-P) \cdots \\
& \quad ~ + \left(  \phi_p(P,R_s) + \phi_{cav}(R_s) \right) ~ \mathcal{H}(P-P_c) \\
& = \frac{1}{2} ~ \left(\frac{a_{\rm inc}R_{\rm inc}}{R_s}\right)^2 \left[ 1 - \frac{P}{P_c} \right] ~ \mathcal{H}(P_c-P) \cdots \\
& \quad ~ + \left( \sqrt{1+a^2(P,R_s)} - 1  + \frac{k_{pe}^2R_s^2}{4}\right) ~ \mathcal{H}(P-P_c) \\
\end{aligned}
\label{eq:nonlinear-eff-potential}
\end{equation}
\begin{equation}
\begin{aligned}
&\frac{d^2 R_s(z)}{dz^2} = \frac{R_{\rm inc}^2}{a_{\rm inc}^2Z_R^2} \times \cdots \\
& \left\{ \frac{1}{R_s}\left(\frac{a_{\rm inc}R_{\rm inc}}{R_s}\right)^2 \left[ 1 - \frac{P}{P_c} \right] ~ \mathcal{H}(P_c-P) ~ + \right. \cdots \\
& \left. \frac{1}{2} \left[ \frac{a^2(P,R_s)}{\sqrt{1+a^2(P,R_s)} ~ (R_s/4)} - k_{pe}^2 R_s \right] ~ \mathcal{H}(P-P_c) \right\}
\end{aligned}
\label{eq:nonlinear-envelope-eq}
\end{equation}

Note that the radial mode of the laser spot over its evolution is assumed to remain $\rm TEM_{00}$ and thus,
\begin{equation}
\begin{aligned}
a^2(r,z) & = \frac{a_{inc}^2 R_{inc}^2}{R_s^2(z)} ~  e^{-2r^2/R_s(z)^2}\\
\frac{\partial a^2(r,z)}{\partial r} \Bigr\rvert_{r=R_s} & = - \frac{1}{R_s/4} ~ a^2(r=R_s,z) 
\end{aligned}
\label{eq:TEM0-mode-pond-force}
\end{equation}

From eq.\ref{eq:nonlinear-envelope-eq}, the matched spot-size, $R^m_{\rm inc}$ is inferred to be a critical point. It is the critical incident spot-size where the ponderomotive force that pushes the envelope out equals to the electrostatic ionic potential that pulls it in and is found to be the same as the bubble radius matching condition in eq.\ref{eq:radial-matched-condition} (under the approximation, $a(r=R_s)\simeq a_{\rm inc}$ and $\sqrt{1+a_{\rm inc}^2}\simeq a_{\rm inc}$).
\begin{equation}
\begin{aligned}
R^m_{\rm inc}=w_0\simeq W_0=2\sqrt{a_{\rm inc}} ~ k_{pe}^{-1}\equiv R_{bubble}
\end{aligned}
\label{eq:spot-size-matched-condition}
\end{equation}

There is an initial radial \enquote{velocity} of the envelope for any value of $R_{\rm inc}\neq R^m_{\rm inc}$. For $R_{\rm inc}>R^m_{\rm inc}$, the ion electrostatic force is dominant and the envelope initially develops a negative radial velocity. Whereas in the opposite limit, $R_{\rm inc}<R^m_{\rm inc}$ the ponderomotive force dominates and the envelope gains a positive initial radial velocity. Here the negative initial velocity condition is primarily studied due its established experimental optimality.

Note that the local changes in group velocity and wavelength within the laser envelope due to local electron density variations within the frame of the laser pulse are not accounted for in this model. Similarly, laser energy depletion is not accounted. Using PIC simulations of the strongly mismatched regime, these effects are shown to become quite important to the envelope behavior.
\begin{figure}[!htb]
   \includegraphics*[width=\columnwidth]{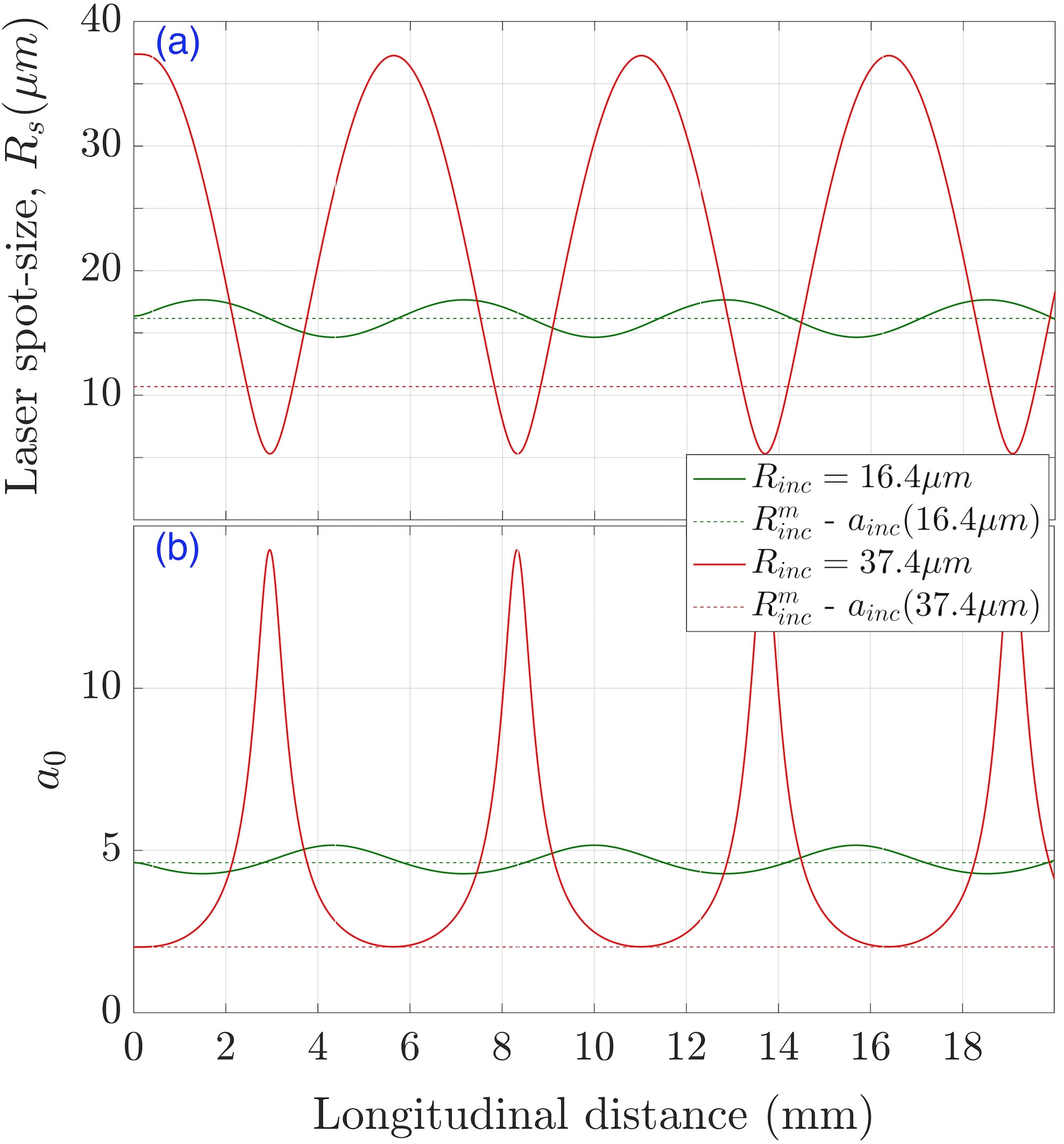}
   \caption{Effect of the mismatch on laser envelope oscillations using eq.\ref{eq:nonlinear-envelope-eq} to compare $R_{\rm inc}=16.4{\rm\mu m}$ and $R_{\rm inc}=37.4{\rm\mu m}$ at $n_0=2\times10^{18}\rm{cm^{-3}}$. (a) envelope spot-size $R_s(z)$ with longitudinal distance, (b) corresponding $a_0$ evolution. The dashed horizontal lines show the matched spot-size, $R^m_{\rm inc}$ at the corresponding incident $a_{\rm inc}$. }
   \label{Fig1:nonlinear-envelope-eq-SOL}
\end{figure}

\begin{figure}[!htb]
   \includegraphics*[width=\columnwidth]{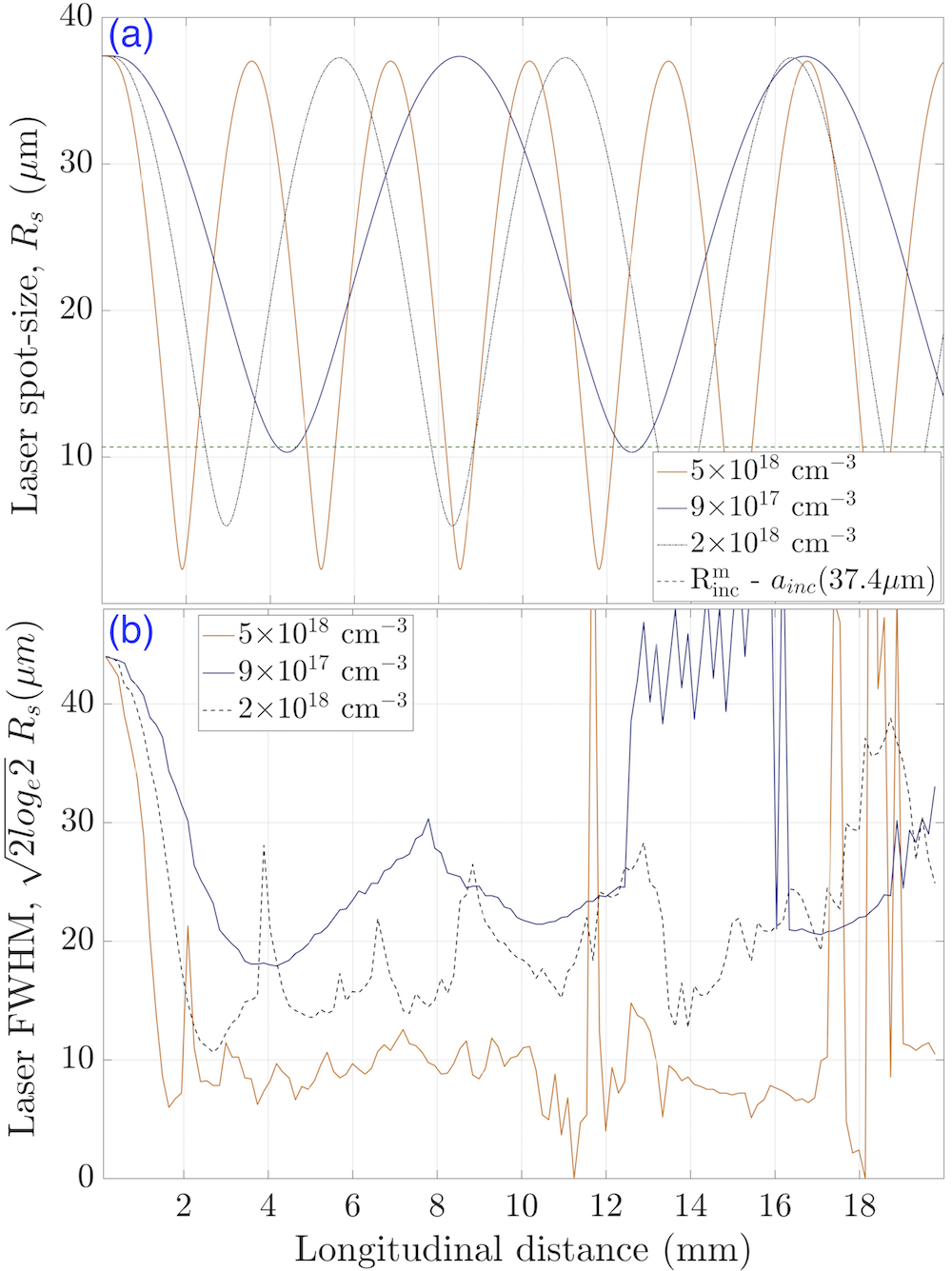}
   \caption{Laser envelope evolution using eq.\ref{eq:nonlinear-envelope-eq} to compare different densities in (a) and envelope evolution from PIC simulations (described in section \ref{sec:pic-simulations}) in (b). This shows the effect of the variation of the degree of mismatch due to the variation of plasma density for $R_{\rm inc}=37.4{\rm\mu m}$ and the other laser parameters the same.}
   \label{Fig2:nonlinear-envelope-eq-SOL-density}
\end{figure}

Numerical solutions of eq.\ref{eq:nonlinear-envelope-eq} are presented in Fig.\ref{Fig1:nonlinear-envelope-eq-SOL} for laser energy of $\mathcal{E}_L=10\rm{J}$ and intensity full-width-at-half-of-maximum (FWHM) pulse length $\tau_p=49\text{fs}$ at electron density,  $n_0=2\times 10^{18} {\rm cm^{-3}}$. The plasma is initialized with a $\rm 500\mu m$ rising plasma density ramp from vacuum to $n_0$ to model experimental conditions and for consistency with PIC simulations. Two different incident spot-sizes, $R_{\rm inc} = w_0 ={\rm 16.4}\rm \mu m$ and $\rm 37.4 \mu m$ (with experimentally relevant intensity FWHM spot-sizes of 19.25$\rm\mu m$ and 44$\rm\mu m$ respectively \cite{Poder-2016}\cite{Kneip-2009}) and corresponding $a_0 \sim$ 4.6 and 2.0 are respectively compared. In Fig.\ref{Fig2:nonlinear-envelope-eq-SOL-density}, the initial focal spot-size is fixed at $R_{\rm inc} = w_0 = 37.4\rm \mu m$ and three different densities are compared.

In Fig.\ref{Fig1:nonlinear-envelope-eq-SOL}(a) spot-size evolution is compared for $R_{\rm inc}=$ 16.4$\rm \mu m$, matched at $n_0=2\times 10^{18} {\rm cm^{-3}}$ for $\mathcal{E}_L=10J$ (shown as dashed line) and strongly mismatched (by a factor of 4) $R_{\rm inc}=$37.4$\rm \mu m$. It is evident that with a high degree of mismatch at $R_{\rm inc}=$37.4$\rm \mu m$, the envelope oscillations become asymmetric whereas they have a small amplitude sinusoidal evolution at $R_{\rm inc}=$ 16.4$\rm \mu m$. It is also seen that in the strongly mismatched regime $a_0$ in Fig.\ref{Fig1:nonlinear-envelope-eq-SOL}(b) sharply rises to many times its incident value in the asymmetric radial squeeze phase. This is a critical result, as it implies rapid variation in the laser-driven plasma wave properties. Note that, $R_{\rm inc}=$37.4$\rm \mu m$ is matched at $n_0=1.5\times 10^{17} {\rm cm^{-3}}$. 

The variation of envelope oscillations over $\rm n_0 = 0.9, 2, 5 \times 10^{18} cm^{-3}$ for a fixed incident spot-size is in Fig.\ref{Fig2:nonlinear-envelope-eq-SOL-density}(a). The oscillation wavelength and the minima of the spot-size in the squeeze phase of the oscillation becomes smaller at higher densities (higher degree of mismatch) while the asymmetry in the oscillations increases.

The nonlinear envelope equation also compares well with the PIC simulations, as shown in Fig.\ref{Fig2:nonlinear-envelope-eq-SOL-density}(b). The details of PIC simulations are in section \ref{sec:pic-simulations}. From simulations at $n_0 = 2\times10^{18} {\rm cm^{-3}}$ with an incident spot-size of $w_0=37.4{\rm \mu m}$, the first radial squeeze minima is at around 3mm, as shown in the numerical solutions of eq.\ref{eq:nonlinear-envelope-eq}. The model also correctly predicts the trend of wavelength of envelope evolution and its minima over a range of densities. However, the envelope behavior changes after the first radial squeeze as seen from the PIC simulation results. Note the PIC snapshots are spaced by 250 fs.

Similar asymmetric envelope oscillation behavior has been reported earlier in \cite{Ehrlich-1996} for sub-critical laser power ($P<P_c$) in a channel-guided mismatched regime, where the incident spot-size is not matched to the matched spot-size for channel-guiding.

It is important to point out that as the radial envelope squeeze phases become shortened and the change in laser envelope radius and $a_0$ becomes more rapid, in simulations it is important to more carefully resolve the radial dimension. In comparison, the matched-regime simulations set a weaker constraint on the resolution of the radial dimensions. In Fig.\ref{Fig1:nonlinear-envelope-eq-SOL}(b) at $n_0=2\times 10^{18} {\rm cm^{-3}}$ the value of $a_0$ varies around its peak over distances which are of the order of $\rm 100 - 200\mu m$. Similarly, in Fig.\ref{Fig2:nonlinear-envelope-eq-SOL-density}(b), the change in radial size is over a few plasma wavelengths. 

In consideration of this important change in the radial envelope dynamics fully resolved 2.5D PIC simulations are thus used instead of transversely under-resolved 3D simulations. Comparison against a well-resolved 3D PIC simulation is used to validate the methodology used in the 2.5D PIC simulations.

This also opens up the case for an \enquote{optical plasma lens}. If the plasma-based focal spot squeezing process is experimentally confirmed to result in a higher focal-spot quality with lesser aberrations compared to vacuum optics, such a lens is quite attractive. From PIC simulations it is observed that the energy loss of the laser over the first squeeze phase is relatively small. This allows the possibility of a mode quality and energy trade-off. This is quite similar but operates based on different physical mechanisms compared to a \enquote{beam plasma lens} \cite{chen-lens-1986}.

\section{Adjusted-$a_0$ model}
\label{sec:adjusted-a0-model}

An \enquote{adjusted-$a_0$ model} is here introduced to account for the mis-match within eq.\ref{eq:radial-matched-condition-energy} and provides a good agreement with experimental observations. This model assumes that the entire laser pulse energy launched at the entrance of the plasma is coupled into the plasma and is then squeezed down to the matched spot-size corresponding to the launched $a_0$. This will therefore increase the $a_0$ by the factor $\Gamma = W_0 / w_0(\text{matched})$ upon the culmination of the squeezing process for a radially symmetric focal spot. Using these heuristic arguments, energy gain in the {\it adjusted-$a_0$ model} is in eq.\ref{eq:energy-strong-mismatch}. F is the optical F-number of the focal spot, $F=\frac{\pi w_0}{2\lambda_0}$. It is related to the focusing optics F-number ($F=f/D$, $f$ is the focal length and $D$ is the aperture of the focusing parabola).
\begin{equation}
\begin{aligned}
\Delta \mathcal{E}_{\text{adj.}}[m_ec^2] & = \frac{2\pi}{3} ~ \sqrt{a_{\rm0-inc}} ~ \frac{W_{0}}{\lambda_0} ~ \sqrt{\frac{n_c}{n_0}} \\
& \simeq 2.5 ~ \sqrt{a_{\rm 0-inc}} ~ \sqrt{\frac{n_c}{n_0}} ~ \text{F}  \\
{\rm circ. : ~} a_0 (\text{adj.}) & = a_{\rm 0-inc} ~ \Gamma = a_{\rm 0-inc} ~ \frac{W_{\rm 0}}{w_{0-m}} \\
{\rm ellip. : ~} a_0 (\text{adj.}) & = a_{\rm 0-inc} ~ \sqrt{ \left(\frac{W_{\rm 0-1}}{w_{\rm 0-m}}\right) \left(\frac{W_{\rm 0-2}}{w_{\rm 0-m}}\right) }
\end{aligned}
\label{eq:energy-strong-mismatch}
\end{equation}

Using eq.\ref{eq:energy-strong-mismatch} the results of many ground-breaking experiments are much better explained where significantly higher energies were obtained in comparison with the predictions of the matched regime model eq.\ref{eq:radial-matched-condition-energy}. For the experiment in \cite{Mangles-2004} eq.\ref{eq:energy-strong-mismatch} predicts peak energy gain of 155 MeV whereas the observed spectral peak was at 70 MeV. From 3D PIC simulations, energies as high as 200MeV are obtained (see Supplementary Material). Similarly, for the experiments in \cite{Faure-2004} eq.\ref{eq:energy-strong-mismatch} predicts a peak energy of 324 MeV whereas the observed spectral peak was at 175 MeV. Note that the lower than theoretically predicted accelerated beam energies are expected in experiments due to experimental factors such as inferior mode quality, $\mathcal{M}^2\gg1$, of the laser focal spot.

This work focusses on GeV-scale acceleration results due to their relevance to current experimental frontiers of the field of LPAs. Experimental data from \cite{Poder-2016} ($\simeq$ 2GeV beam) and \cite{Kneip-2009} ($\simeq$ 1GeV beam) are used to further detail the mismatched regime.
\begin{table}[!htb]
\setlength{\tabcolsep}{20pt}
\begin{center}
\captionsetup{belowskip=5pt,aboveskip=2pt}
\caption{Laser, Plasma and $e^-$ beam parameters in \cite{Poder-2016}}
\label{table:Gemini-f40-parameters}
\begin{tabular}{ l  l }
\\ [-3.0ex]\hline 
\hline \\[-2.4ex]
 	$\rm{\mathcal{E}_L}$, FWHM-$\mathlarger{\mathlarger{\tau_p}}$ 				& $\simeq$ 10 J, 49 fs \\
	$P_L$																		& $\simeq$ 200 TW \\
  	$W_{0-y}$ (y-axis) 				 											& 37.4 $\mu m$  \\
	$W_{0-z}$ (z-axis) 				 											& 44.2 $\mu m$  \\
	${\rm incident} ~ a_0 ~ (a_{\rm inc})$												& $\simeq 1.9$ \\
 	$\Delta\mathcal{E}_{\rm{peak}}$												& 2.2 GeV \\
	$n_0$ (peak energy)			    											& $2-3 \times 10^{18} ~ \rm{cm^{-3}}$ \\
	$P_c$																		& $18.3 - 12.2$ TW \\
\hline	 
\end{tabular}
\end{center}
\end{table}%

\begin{figure}[!htb]
   \includegraphics*[width=\columnwidth]{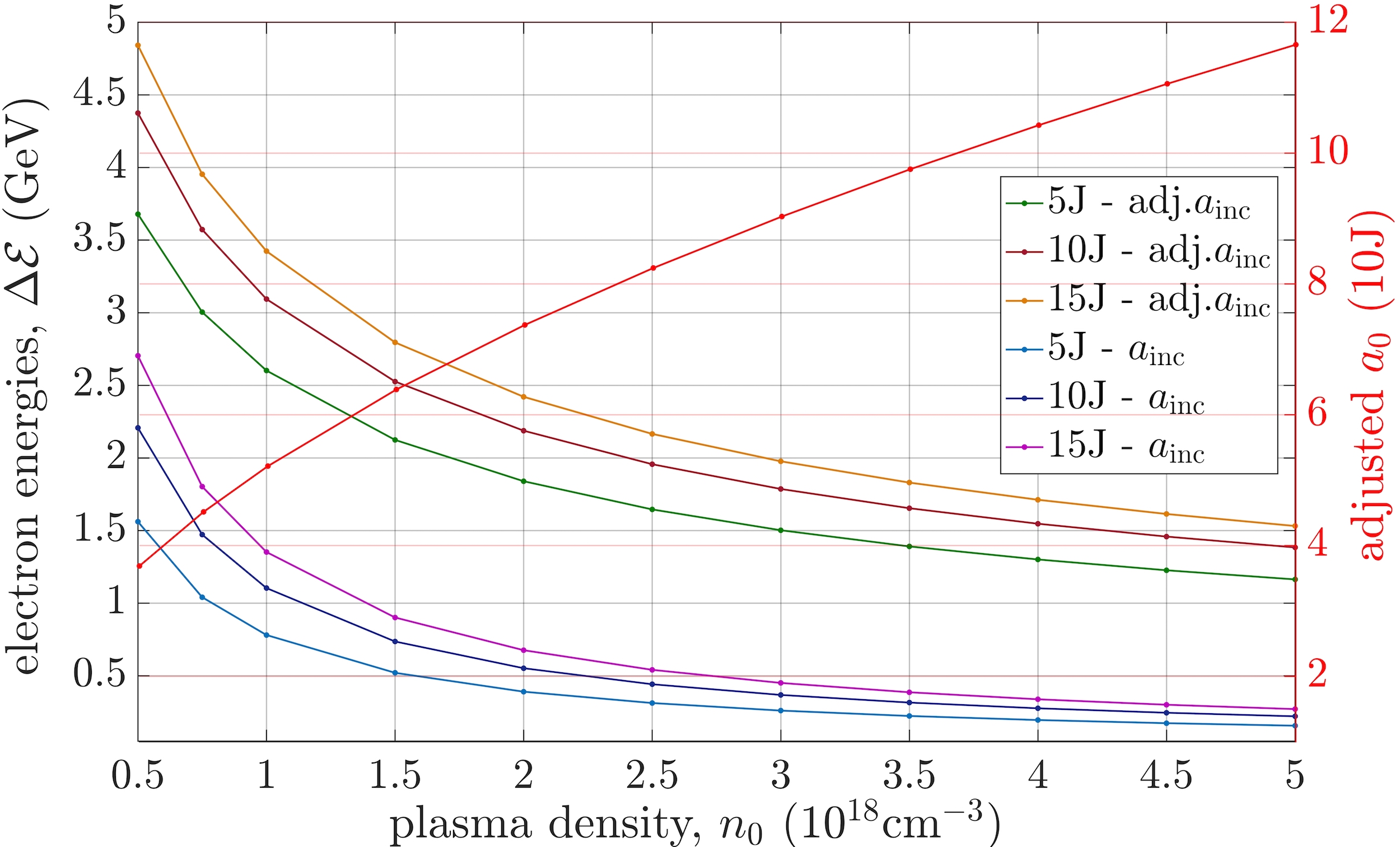}
   \caption{ Comparison of electron beam energies predicted by eq.\ref{eq:radial-matched-condition-energy} \cite{Lu_PRSTAB_2007} to the energies from the adjusted-$a_0$ model for laser energies $\rm{\mathcal{E}_L}$=5, 10 \& 15 J (corresponding to parameters in Table.\ref{table:Gemini-f40-parameters}).}
   \label{Fig:e-beam-energy-scaling}
\end{figure}

The predicted accelerated beam energies from eq.\ref{eq:energy-strong-mismatch} in comparison with eq.\ref{eq:radial-matched-condition-energy} for laser and plasma parameters of the experiments in \cite{Poder-2016} (in Table.\ref{table:Gemini-f40-parameters}) are shown in Fig.\ref{Fig:e-beam-energy-scaling}. The peak energies predicted by eq.\ref{eq:radial-matched-condition-energy} for $n_0 = 1.5-3 \times 10^{18} ~ \rm{cm^{-3}}$ are $\leq \rm{1GeV}$, whereas the experiments obtained energy $\rm> 2 GeV$. Thus, at lower densities and higher intensities the disagreement between the predictions of the 3D simulation based matched regime model of \cite{Lu_PRSTAB_2007} and the experiments grows. 

This significant disagreement between the electron energies predicted by eq.\ref{eq:radial-matched-condition-energy} and the experimentally obtained energies, is because the laser-plasma acceleration process in \cite{Poder-2016} corresponds to a \enquote{strongly mismatched regime}. This is evident from eq.\ref{eq:radial-matched-condition}, the matched $w_0$ for this interaction at the incident $a_0$ is $\rm 10.3\mu m$ and for incident laser energy is $\rm 16.4\mu m$ at $n_0 = 2 \times 10^{18} ~ \rm{cm^{-3}}$. In contrast, the launched elliptical laser spot-size has its minor-axis waist-size of $37.4 ~\mu m$ (a factor of 4 mismatch).

It is important to note that in \cite{Poder-2016}, the matched spot-size is $\rm 16.4\mu m$ for laser energy of 10J. Additionally, the experiments reported at this matched spot-size in \cite{Kneip-2009} with 10J laser energy only reported 100MeV energy gain at $n_0<5\times10^{18}cm^{-3}$.

The adjusted-$a_0$ model in eq.\ref{eq:energy-strong-mismatch} is used to calculate the expected electron energies for parameters of the experiments in \cite{Poder-2016} (in Table.\ref{table:Gemini-f40-parameters}). In these experiments a linearly polarized laser with $a_0 \simeq 1.9$ and spot-size $\rm W_0 = 37.4 \mu m$ is incident on a plasma with density $\rm n_0 = 2 \times 10^{18} \rm{cm^{-3}}$. The electron beam energy expected from eq.\ref{eq:radial-matched-condition-energy} is $\rm < 1 ~ GeV$ whereas from experiments energy gain $\Delta \mathcal{E} = \rm{2.2 ~ GeV}$. Using the adjusted-$a_0$ model eq.\ref{eq:energy-strong-mismatch} the predicted energy gain is $\rm\Delta \mathcal{E} ~ [a_0(adj) = 7.4] = \rm{2.2 ~ GeV}$.

For the parameters in Table.\ref{table:Gemini-f40-parameters}, the value of average accelerating plasma field from \cite{Lu_PRSTAB_2007} is $\rm \langle E_{\rm acc}\rangle(a_0=1.9) = 0.5 \sqrt{a_0} ~ m_ec\omega_{pe}e^{-1} = 93.7 ~ GVm^{-1}$ but with the adjusted-$a_0$ model it is $\rm \langle E_{\rm acc}(a_0[adj.]=7.4) \rangle = 185 ~ GVm^{-1}$. Note that, at $2\times10^{18} \rm{cm^{-3}}$, the wave-breaking field is $\rm{E_{wb}} \equiv m_ec\omega_{pe}e^{-1} = 136 ~ GVm^{-1}$. The adjusted-$a_0$ model accurately predicts even in this case, because in the concerned experiments the peak beam energy of 2.2 GeV is gained over a total of 13mm (length up to the injection point in not accounted). This gives an average acceleration gradient of $\simeq 170 ~ \rm GVm^{-1}$. It is found that that the peak plasma fields in this regime are of the order of $a_0(\rm{adj.}) \times m_ec\omega_{pe}e^{-1} = 1006 ~ GVm^{-1}$. Peak fields of $\rm 800 ~ GVm^{-1}$ are observed in simulations (see Fig.\ref{Fig4:plasma-size-field-evolution-f40}(b)) where data is only available in increments of 250 fs.

Not surprisingly, an agreement with the energy-gain predictions of the adjusted-$a_0$ model is also obtained for experiments reported in \cite{Kneip-2009}. At $5.5 \times 10^{18} ~ \rm{cm^{-3}}$ with $a_0=3.9$, the matched $w_0$ is $8.95\mu m$ whereas the launched $w_0=19\mu m$. The energy expected from the matched regime formula in eq.\ref{eq:radial-matched-condition-energy} is 510 MeV. The adjusted $a_0$-model predicts a beam energy of $957 ~ \rm{MeV}$ with $a_0 \rm{(adj.)}=8.3$, whereas a spectral peak was experimentally observed at $\rm \simeq 800 ~ MeV$.

In the sections that follow PIC simulations are used to show that the heuristic arguments that lead to this good agreement have their basis in the atypical laser envelope evolution. The laser energy in the incident spot is shown in simulations to squeeze down to the matched spot-size over a wide range of densities as discussed in section.\ref{sec:nonlinear-envelope-equation}. The laser waist is shown to subsequently remain close to the matched spot-size. 

But, not all the processes that play a role in electron acceleration to high energies are fully explained with the adjusted-$a_0$ model. The important observations of optimal densities range $\rm n_0 = 1.5 - 3 \times 10^{18} ~ \rm{cm^{-3}}$ for the experiments in \cite{Poder-2016} and a similar optimal range $\rm n_0 = 5-7 \times 10^{18} ~ \rm{cm^{-3}}$ in \cite{Kneip-2009} raises many questions. It is clear from above that the {\it adjusted-$a_0$ model} is not applicable over a broad range of densities.

\section{Multi-dimensional PIC Simulations}
\label{sec:pic-simulations}
In this section 3D and 2.5D PIC simulation (two spatial and three velocity dimensions) results with parameters from accessible experimental data are presented using the \textsc{epoch} code \cite{EPOCH-paper}. 

An analysis of the simulations reveals the processes that underlie the acceleration mechanism in the strongly mismatched regime and shows the complex laser-plasma interaction dynamics. It also shows that these mechanisms significantly differ from the matched regime. 

The {\it general applicability of the strongly mismatched regime} model is established by proof of the equivalence of physical mechanisms in two different experiments that use the strongly mismatched regime \cite{Mangles-2004} and \cite{Poder-2016} with entirely different laser and plasma parameters. The experiments in \cite{Mangles-2004} are modeled with 3D and 2.5D PIC simulations whereas in \cite{Poder-2016} only with 2.5D simulations.

The {\it validity of the methodology} of using $\rm 2 \times a_0$ in 2.5D simulations for equivalence to 3D PIC simulations and thus to the experiments is proved by the comparison of these simulation for \cite{Mangles-2004}. From the movies in Supplementary Materials that compare the evolution of electron density, laser field and plasma field an excellent agreement is found between $\rm 2 \times a_0$-2.5D and 3D PIC simulations. A good agreement is also found in the evolution of the beam energy spectra for 2.5D and 3D simulations (movies provided). Further validation of the equivalence of the 2.5D and 3D PIC simulations is established by a good agreement between the evolution of the bubble size and the laser pulse length also in the Supplementary Materials. In $\rm 2 \times a_0$-2.5D PIC simulation the initially boosted vacuum-$a_0$ primarily accounts for the squeeze-phase peak plasma-$a_0$ expected from the non-linear envelope equation.

The two signature processes of the strongly mismatched regime - strong optical shock and bubble elongation, are clearly evident in 3D and 2.5D simulations of the experiments in \cite{Mangles-2004} and \cite{Poder-2016}. These are shown to agree with the location of the squeeze phases predicted form the non-linear envelope equation in eq.\ref{eq:nonlinear-envelope-eq} (for \cite{Mangles-2004} shown in Supplementary Material). The correspondence of the slicing of the laser pulse that drives the optical shock in 3D and 2.5D PIC simulations in \cite{Mangles-2004} is shown using Wigner-Ville transform snapshots in Supplementary Materials. The elongation of bubble is also evident in both these simulations from the movies.

The simulations are setup in a moving simulation box which tracks the laser pulse at its unperturbed group velocity, is used. A linearly polarized laser with a Gaussian envelope is initialized such that it entirely enters the box, before the box starts moving. Absorbing boundary conditions are used for both the fields and particles. The laser pulse is incident from the left boundary (using a laser boundary condition) and propagates in $\rm50\mu m$ of free-space before it focusses onto the plasma with a spot-size equal to the minor axis of the elliptical focal spot before the box starts to move. 

In 2.5D simulations a cartesian grid is used with 25 cells per laser wavelength ($\lambda_0$) in the longitudinal direction and 15 cells per laser wavelength in the transverse. In 3D simulations the longitudinal direction is resolved with 22.5 cells per laser wavelength and the two transverse direction with 7.5 cells per laser wavelength. The 2.5D simulations are initialized with 4 particles per cell and 3D with 1 particle per cell.

A gas-jet is simulated with a linear density gradient of $\rm50\mu m$ before the homogeneous plasma whereas the gas-cell has a $\rm500\mu m$ linear density gradient to mimic the measured electron density profile.

\subsection{Dynamics of Laser-Plasma Interaction}
\label{sub-sec:analysis-PIC-simulation}
\begin{figure}[!ht]
   \includegraphics*[width=\columnwidth]{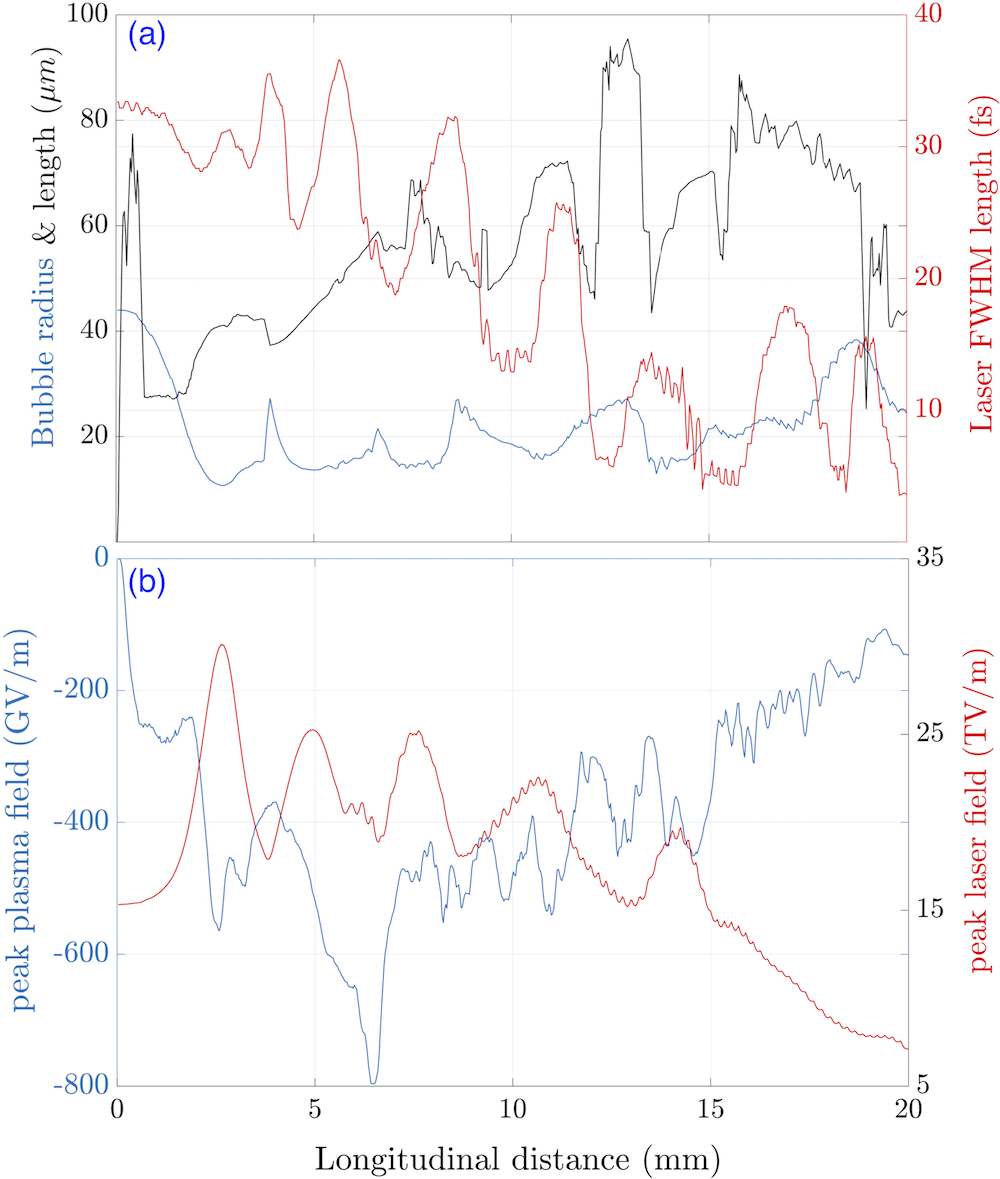}
   \caption{Results from 2.5D PIC simulation at $\rm n_0=2\times 10^{18} cm^{-3}$ (a) the evolution of the bubble radial (blue), length (black) and laser pulse length from E-field (red) (b) evolution of peak laser field (red) and plasma longitudinal field (blue) with propagation distance using parameters in table \ref{table:Gemini-f40-parameters}. }
   \label{Fig4:plasma-size-field-evolution-f40}
\end{figure}

The laser energy evolution does not exhibit any correlation with the electron beam energy or possess any specific signature of the laser-plasma interaction process in the mismatched regime. Interestingly, the energy loss over the first \enquote{squeeze}-phase is relatively small which allows this regime to be useful as an efficient {\it optical plasma lens} and the {\it adjusted-$a_0$} model to be valid. 

An analysis of the evolution of the laser-plasma dimensions and fields as shown in Fig.\ref{Fig4:plasma-size-field-evolution-f40} is much more instructive. Fig.\ref{Fig4:plasma-size-field-evolution-f40}(a) shows the evolution of dimensions and Fig.\ref{Fig4:plasma-size-field-evolution-f40}(b) the evolution of fields for $n_0 = 2\times10^{18} \rm{cm^{-3}}$. Enumerated below are several laser-plasma effects of interest that are inferred by the study of this evolution.
\begin{enumerate}[nolistsep,label=(\roman*)]
\item The laser pulse intensity-FWHM waist size (in blue in \ref{Fig4:plasma-size-field-evolution-f40}(a)) launched at 44$\mu m$ is squeezed down to a minimum spot-size of $\simeq 10\mu m$ in 10 ps ($\simeq$3mm) in good agreement with eq.\ref{eq:nonlinear-envelope-eq} and Fig.\ref{Fig1:nonlinear-envelope-eq-SOL}. This process of the initial focal spot nearly squeezing down to the incident-intensity matched spot-size occurs over a wide range of densities.
\item The laser pulse {\it radial envelope} oscillates due to the strong initial mismatch. However, the spot-size remains close to the matched spot-size which corresponds to the squeezing down of the laser energy to around $\rm 15 \mu m$. The maximum radial excursions are less than half the launched spot-size. More importantly these are all precursors to the successive triggering of a state of \enquote{strong optical shock}, which inhibits free radial expansion predicted by the nonlinear envelope equation, eq.\ref{eq:nonlinear-envelope-eq}. This radial confinement explains the agreement of the experiments in the strongly mismatched regime to the adjusted-$a_0$ model, which is based upon the laser energy squeezing to the matched spot-size.
\item The laser pulse longitudinal or temporal envelope undergoes events of \enquote{catastrophic collapses}. This is inferred from the evolution of field-FWHM time duration of the laser pulse (in red) over time. There are about 4 such events in  \ref{Fig4:plasma-size-field-evolution-f40}(a) around 4.2mm, 6.3mm, 8.7mm and 12mm. A rapid collapse of the laser time-FWHM indicates the triggering of a strong \enquote{optical shock} due to slicing of the laser. This leads to a sharp laser-front edge.
\item This laser slicing effect is observed to correspond with a rapid increase in the bubble length. The length of the bubble is initially equal to its radius. But, as the laser radial envelope squeezes the bubble length rapidly increases while its radius remains almost constant, as seen from the comparison of the blue and the black curve (in \ref{Fig4:plasma-size-field-evolution-f40}(a)). This is due to the optical shock driven rapid elongation of the bubble.
\item The triggering of {\it optical shock} state and the excitation of a rapidly {\it bubble elongation} directly corresponds to the injection of electrons in the back of the lengthened yet radially stable bubble. 
\end{enumerate}

The laser-plasma interactions effects that underlie the acceleration mechanism are also reflected in the evolution of the laser and plasma fields in Fig.\ref{Fig4:plasma-size-field-evolution-f40}(b). It is observed that the peak plasma field occurs when the laser pulse temporal field-FWHM starts to undergo a sudden collapse. The triggering of a strong optical shock drives a rapid bubble elongation with peak plasma field of $\simeq -800 ~ \text{GV/m}$ at around 6.3mm. The highest energy bunches are injected as the bubble rapidly elongates in response to the rapid increase in longitudinal ponderomotive force from the steep rise in the intensity at the head of the optical shock.
\begin{equation}
\begin{aligned}
& \mathcal{E}_{quiv}^e(x,r) \propto I(x,r)\lambda_0^2(x,r) \\
\text{Spherical bubble :} & \nabla_{\parallel} \mathcal{E}_{quiv}^e(x,r) \simeq \nabla_r \mathcal{E}_{quiv}^e(x,r)  \\
\text{Elongated bubble :} & \nabla_{\parallel} \mathcal{E}_{quiv}^e(x,r) \gg \nabla_r \mathcal{E}_{quiv}^e(x,r)
\end{aligned}
\label{eq:dis-balanced-ponderomotive-force}
\end{equation}
\begin{figure}[!htb]
   \includegraphics*[width=\columnwidth]{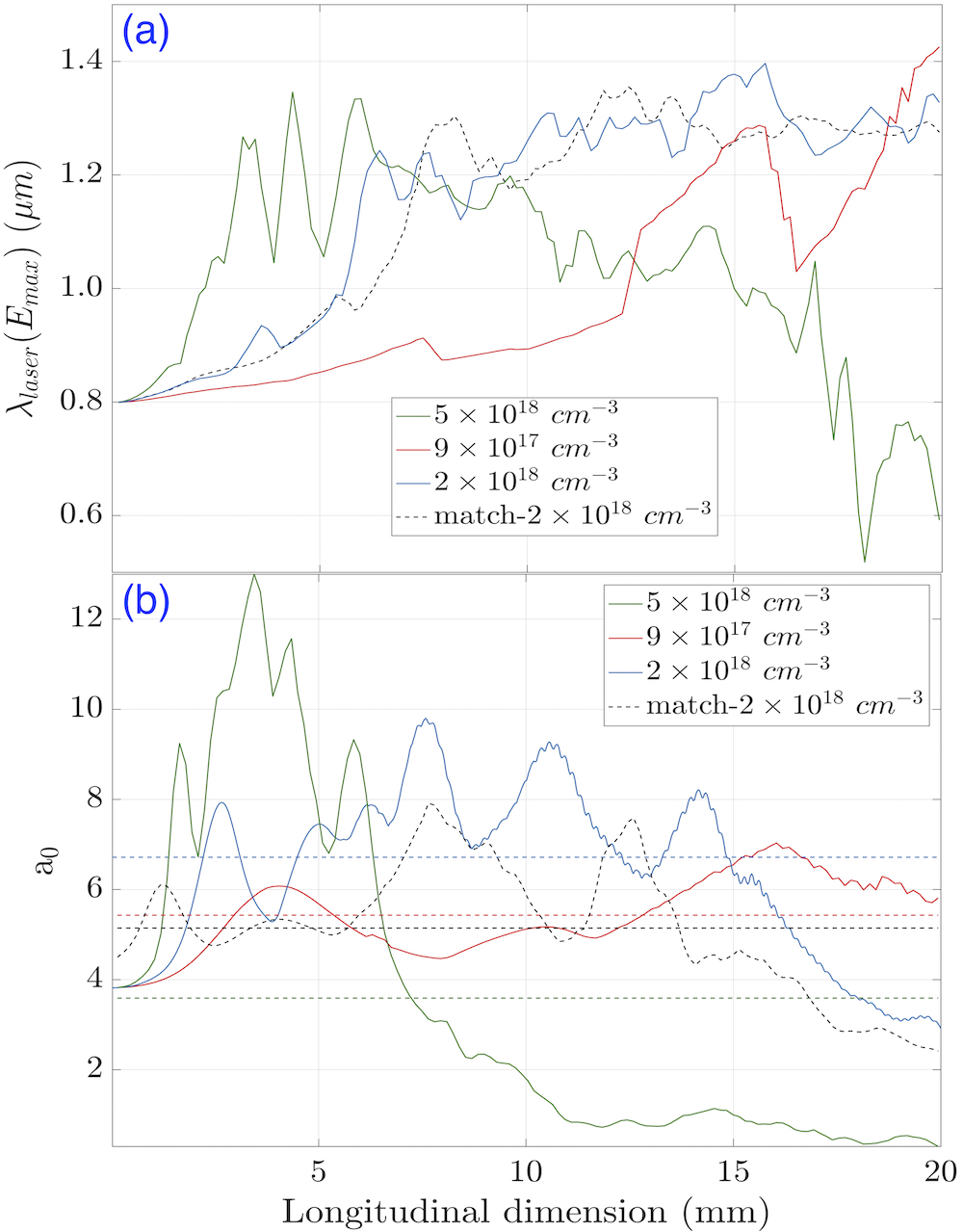}
   \caption{Results from 2.5D PIC simulation that compare different densities (a) the evolution of laser wavelength at the location of the peak laser field amplitude from averaged Wigner-Ville transform and (b) the evolution of peak laser normalized vector potential obtained with $a_0 \propto \sqrt{I\lambda^2}$. }
   \label{Fig5:laser-wavelength-a0-evolution-f40}
\end{figure}

This is a novel injection mechanism due to the imbalance between the longitudinal and radial ponderomotive force, a condition represented in eq.\ref{eq:dis-balanced-ponderomotive-force}, where $\mathcal{E}_{quiv}^e(x,r)$ is the energy of quiver motion in the laser field. The ponderomotively driven electrons have different longitudinal and radial oscillation periods. The injection occurs due to the lower radial momentum of the electrons pushed by the low-intensity sliced part of the pulse ahead of the shock. In particular, as these electron return to the bubble axis much ahead of the electrons driven by the optical shock they experience the shock-driven bubble fields.

\begin{figure}[!htb]
   \includegraphics*[width=\columnwidth]{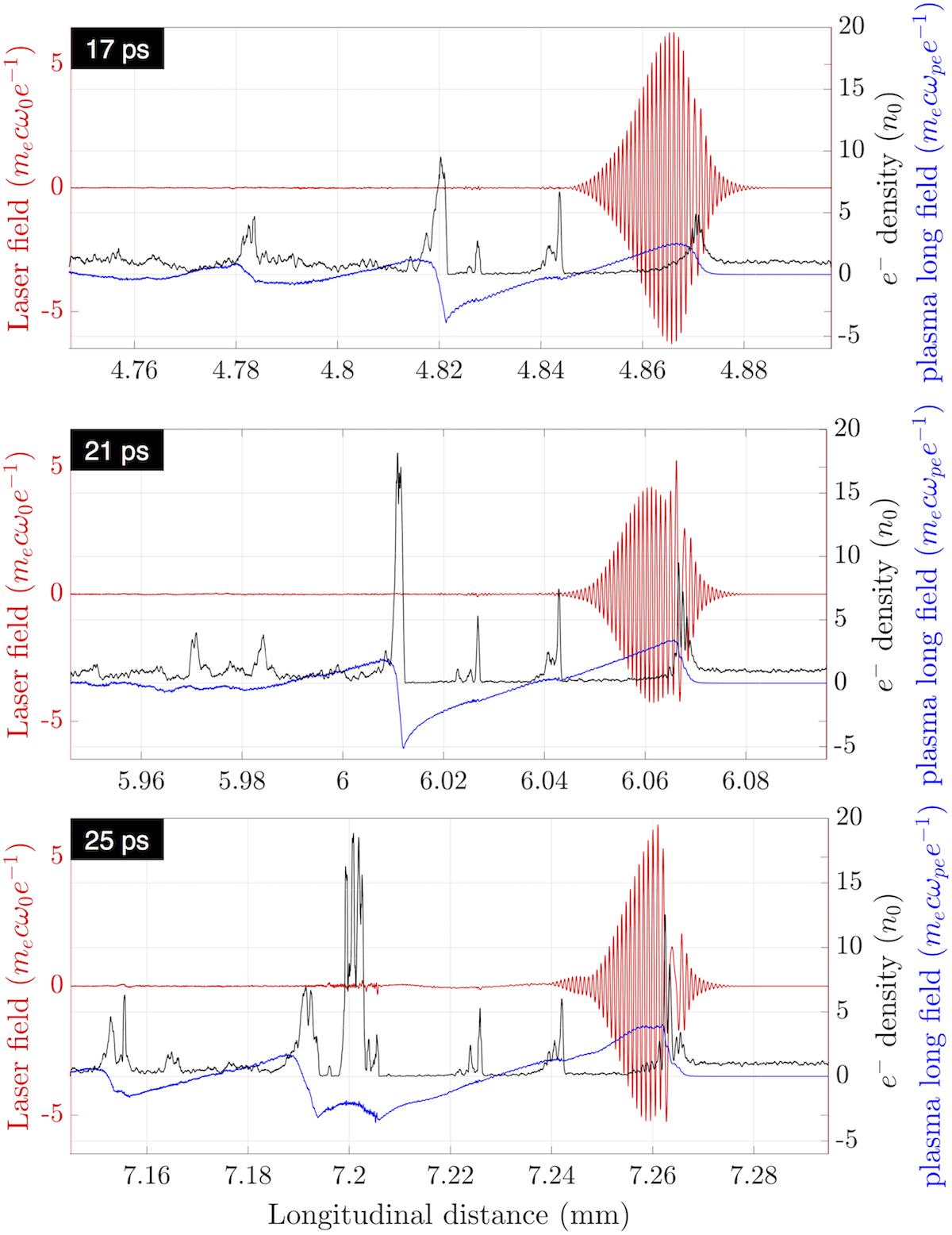}
   \caption{On-axis line-out from 2D PIC simulations for $\rm n_0=2\times10^{18}cm^{-3}$ which shows the formation of the optical shock in the strongly mismatched regime. The on-axis laser transverse electric field is in red. The plasma longitudinal field (blue curve) is used to infer the bubble length which is seen to undergo an increase. The electron density is in black.}
   \label{Fig6:optical-shock-formation}
\end{figure}

At 6.3mm (around 24ps) as seen in Fig.\ref{Fig4:plasma-size-field-evolution-f40}(b) and Fig.\ref{Fig6:optical-shock-formation}, the optical shock is excited by slicing the laser close to the peak laser field and the laser energy is still high enough in its evolution to cause the strongest dis-balance between the radial and longitudinal forces (ponderomotive force evolution is in the Supplementary material).

A clear insight is also developed into the reasons behind an optimal density range in the mismatched regime for the highest energy gain using Fig.\ref{Fig5:laser-wavelength-a0-evolution-f40}. In Fig.\ref{Fig5:laser-wavelength-a0-evolution-f40}(a), the laser wavelength at the maximum laser field is shown for different densities with the optimum at $\rm n_0=2\times10^{18}cm^{-3}$. It is observed that at the lower end of the optimum density range ($\rm n_0 = 9\times 10^{17}cm^{-3}$, in red) the wavelength change is slow and a jump in wavelength which corresponds with a shock occurs only around 12.5mm, where the laser has significantly depleted. On the other hand, at the higher end of the optimum density range ($\rm n_0 = 5\times 10^{18}cm^{-3}$, in green), the triggering of shock occurs multiple times and rapidly depletes the laser pulse. 

This is further demonstrated by the evolution of peak-$a_0$ in Fig.\ref{Fig5:laser-wavelength-a0-evolution-f40}(b) for different densities. At the lower density end of optimum, the value of $a_0$ increases too slowly and for $\rm n_0 = 9\times 10^{17}cm^{-3}$ its average value over 20mm is $\langle a_0 \rangle=5.4$. At the higher density end of optimum, the value of $a_0$ initially increases too rapidly and falls to well-below its initial value before 10mm, with the average over 20mm being $\langle a_0 \rangle=3.6$. For the optimum density at $\rm n_0 = 2\times 10^{18}cm^{-3}$ the average value of $\langle a_0 \rangle=6.7$, which is quite comparable to the value arrived at in the adjusted-$a_0$ model of 7.4.

In the matched regime which is simulated here with $a_0\sim5$ and $w_0=16.4\rm{\mu m}$ (the matched parameters used in the nonlinear envelope equation, eq.\ref{eq:nonlinear-envelope-eq} shows minimum envelope oscillations) the rate of wavelength change is much slower than in the mismatched regime. The average value of $a_0$ in the matched regime over 20mm is much lower in the matched regime at $\rm \langle a_0 \rangle=5.1$ compared to the optimum for the mismatched regime at $\langle a_0 \rangle=6.7$.

Here the wavelength is calculated from the Wigner-Ville transform of the laser-field on-axis line-out. The wavelength distribution function is averaged at each longitudinal location. The value of normalized vector potential $a$ is calculated at each location from its dependence on $I\langle\lambda\rangle^2$ and the peak value is used as $a_0$.

\subsection{Strong optical-shock excitation by laser slicing \\ and Bubble elongation}
\label{sub-sec:strong-optical-shock}
The {\it triggering} of an optical shock excitation is studied from the evolution dynamics of the laser envelope in the strongly mismatched regime. In this regime, due to the oscillations of the laser radial envelope each of the asymmetric envelope-squeeze event drives the laser radial envelope towards the matched spot-size. This causes a large surge in the peak laser field. It is observed that each of the surge in the laser electric fields triggers an {\it optical shock excitation} event.
\begin{equation}
\begin{aligned}
\beta_g = \beta_{\phi-las}^{-1} \simeq ~ & \beta_{g0} ~ \left[ 1 + \frac{1}{2\gamma_{g0}^2\beta_{g0}^2} \left(\langle a_{\perp} \rangle^2 - \frac{\delta n}{n_0}\right) \right] \\
& \beta_{g0} = \sqrt{1 - \frac{\omega_{pe}^2}{\omega_0^2} } , \gamma_{g0} = \frac{\omega_0}{\omega_{pe}} 
\end{aligned}
\label{eq:third-order-dispersion-velocity}
\end{equation}

The third-order perturbative expansion based relation for the laser pulse group velocity \cite{sahai-thesis-2015} in a quasi-static plasma wake with local parameters of plasma ($\delta n(\xi,r)/n_0$) and laser ($a(\xi,r)$) dictates the local group velocity as in eq.\ref{eq:third-order-dispersion-velocity} ($\xi=c\beta_{g0}t-z$, is a coordinate that co-propagates with the laser). This equation is used to treat spatially-localized laser-plasma interaction, because it handles group velocity $\beta_g(\xi,r)$ at each point in space in the co-moving coordinate.

This relation in eq.\ref{eq:third-order-dispersion-velocity} is used to estimate a {\it locally zero group velocity} condition shown in eq.\ref{eq:zero-local-group-velocity}. 
\begin{equation}
\begin{aligned}
\beta_g(\xi,r)  & = 0 \\
\frac{1}{2}\left(\frac{\delta n}{n_0} - \langle a_{\perp} \rangle^2\right) & \simeq \frac{n_c}{n_0}
\end{aligned}
\label{eq:zero-local-group-velocity}
\end{equation}

Though the zero local envelope velocity condition is a mathematical construct because in this work the typical initial plasma density is around $n_c/n_0 \simeq 30$, it does essentially demonstrate that the group velocity of a pulse in different parts of the wake significantly differs. It is quite evident that if $\delta n(\xi,r) \rightarrow n_c$ when $\langle a_{\perp}(\xi,r) \rangle^2\rightarrow 0$, then the local group velocity, $\beta_g(\xi,r)\rightarrow 0$. This implies that a part of the envelope gets slowed down much more in comparison to the rest of the pulse and thus this part of the envelope is lost. This leads to \enquote{slicing of the laser} into two distinct pulses under the conditions above. It should also be noted that the $\beta_g(\xi,r)\rightarrow 0$ implies $\beta_{\phi-las}(\xi,r)\rightarrow\infty$ which means $\lambda_{las}(\xi,r)\rightarrow \infty$. So, an increase in the wavelength in a local region implies a local reduction in the group velocity.

The time evolution of the on-axis laser field from PIC simulations is shown for one such event in Fig.\ref{Fig6:optical-shock-formation} which corresponds to the triggering of an optical shock at 21ps and its formation based on the completion of slicing at 25ps. A rapid increase in the laser wavelength for $\rm n_0=2\times10^{18}cm^{-3}$ is shown with the average wavelength plotted in the Fig.\ref{Fig5:laser-wavelength-a0-evolution-f40}(a). Here around 21ps the laser wavelength has rapidly jumped to $\rm 1.2\mu m$, from the initially launched value of $\rm 0.8\mu m$.

The laser-plasma interaction dynamics that underlies \enquote{laser slicing} is also seen in parameters other than the on-axis dynamics in Fig.\ref{Fig6:optical-shock-formation}. These parameters are the radial intensity-FWHM in Fig.\ref{Fig4:plasma-size-field-evolution-f40}(a), the laser field in Fig.\ref{Fig4:plasma-size-field-evolution-f40}(b), and the laser wavelength in Fig.\ref{Fig5:laser-wavelength-a0-evolution-f40}(a). 

Fig.\ref{Fig6:optical-shock-formation} shows the laser-plasma interaction dynamics in the front of the bubble. It is seen that at 17ps the wavelength in the front of the wake, a region collocated with max-$\delta n/n_0$ (where the longitudinal ponderomotive force is highest) begins to increase. This time also corresponds to a rapid surge in the laser electric field and thus the ponderomotive force rapidly increases. This also leads to an increase in the max-$\delta n/n_0$ at the laser head. At 21ps, in the region of max-$\delta n/n_0$ the wavelength has significantly stretched. This corresponds to a rapid reduction in the local group velocity. At 25ps, the laser envelope is broken into two distinct regions separated by a long wavelength, low group velocity cycle. These laser cycles of long-wavelength low group-velocity lead to the detachment of the head of the laser pulse from it and the triggering of optical shock state. The duration of the persistence of sliced laser is also the time where the laser longitudinal ponderomotive force becomes largely imbalanced with the radial ponderomotive force.

The large imbalance between the longitudinal and the radial ponderomotive force is seen to have a direct effect on the length and the radial envelope of the laser. The bubble length is seen to grow much more than the bubble radius. The bubble elongation driven by the large longitudinal ponderomotive force has a direct effect on the peak longitudinal plasma field which grows to around -800 GV/m at 25ps, as seen in Fig.\ref{Fig4:plasma-size-field-evolution-f40}(b). 

The bubble elongation that follows an optical shock excitation also drives the self-injection of a large amount of charge on to the bubble axis. Because the injection of charge occurs when the laser is in the state of an optical shock, the injected charge experiences much higher peak plasma field and accelerates to peak energies in less than a centimeter. The elongated bubble also has a longer de-phasing length while it lasts in the elongated state.

The {\it laser slicing} process is also observed in the 3D and 2.5D simulations of \cite{Mangles-2004}. The slicing of laser is evident at 2.2ps as seen from the Wigner-Ville transform and the corresponding on-axis line-outs in the Supplementary Material. The bubble also starts to elongate around 2.2ps driven by the optical shock as seen in the evolution movies. This also shows good agreement with the onset of the envelope squeeze phases obtained from the solution of envelope equation eq.\ref{eq:nonlinear-envelope-eq} presented in Supplementary Materials.

\subsection{Properties of the beam injected by \\ strong optical-shock based self-injection}
\label{sub-sec:beam-properties}

\begin{figure}[!htb]
   \includegraphics*[width=\columnwidth]{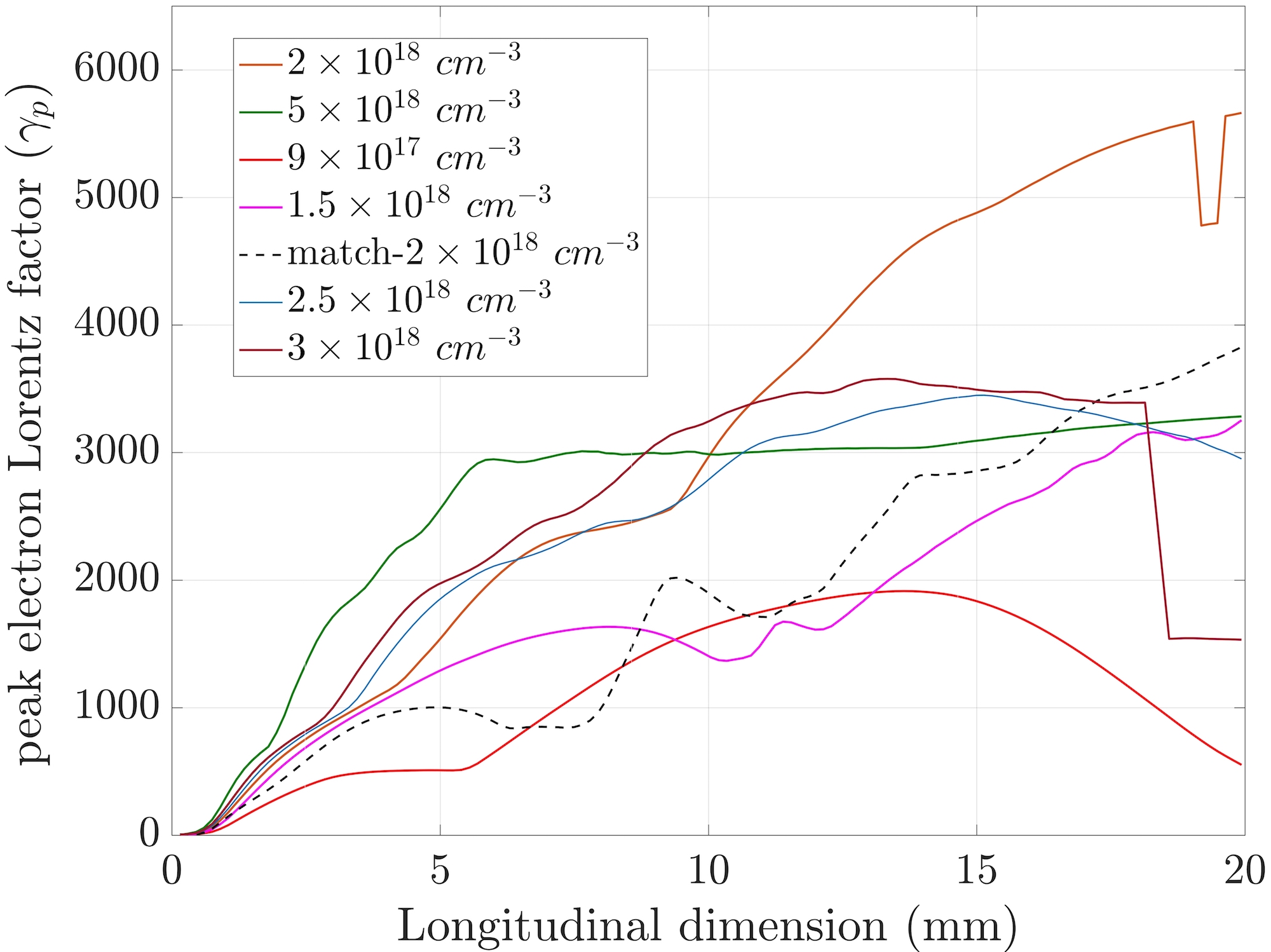}
   \caption{Peak Lorentz factor ($\gamma_p$) evolution which corresponds with the energy gain, $\Delta\mathcal{E}$ compared for various densities and to the matched regime, from 2.5D PIC simulations of the strongly mismatched regime for laser energy $\rm{\mathcal{E}_L}$ = 10J.}
   \label{Fig7:peak-beam-energy-evolution}
\end{figure}

The properties of the accelerated electrons that are injected by the novel strong \enquote{optical-shock} driven \enquote{bubble elongation} extracted from the PIC simulations is presented for parameters in \cite{Poder-2016}. In the strongly mismatched regime, the laser and the bubble properties critically depend upon the plasma density and the degree of mismatch and thus the beam properties reflect this.

\begin{figure}[!htb]
   \includegraphics*[width=\columnwidth]{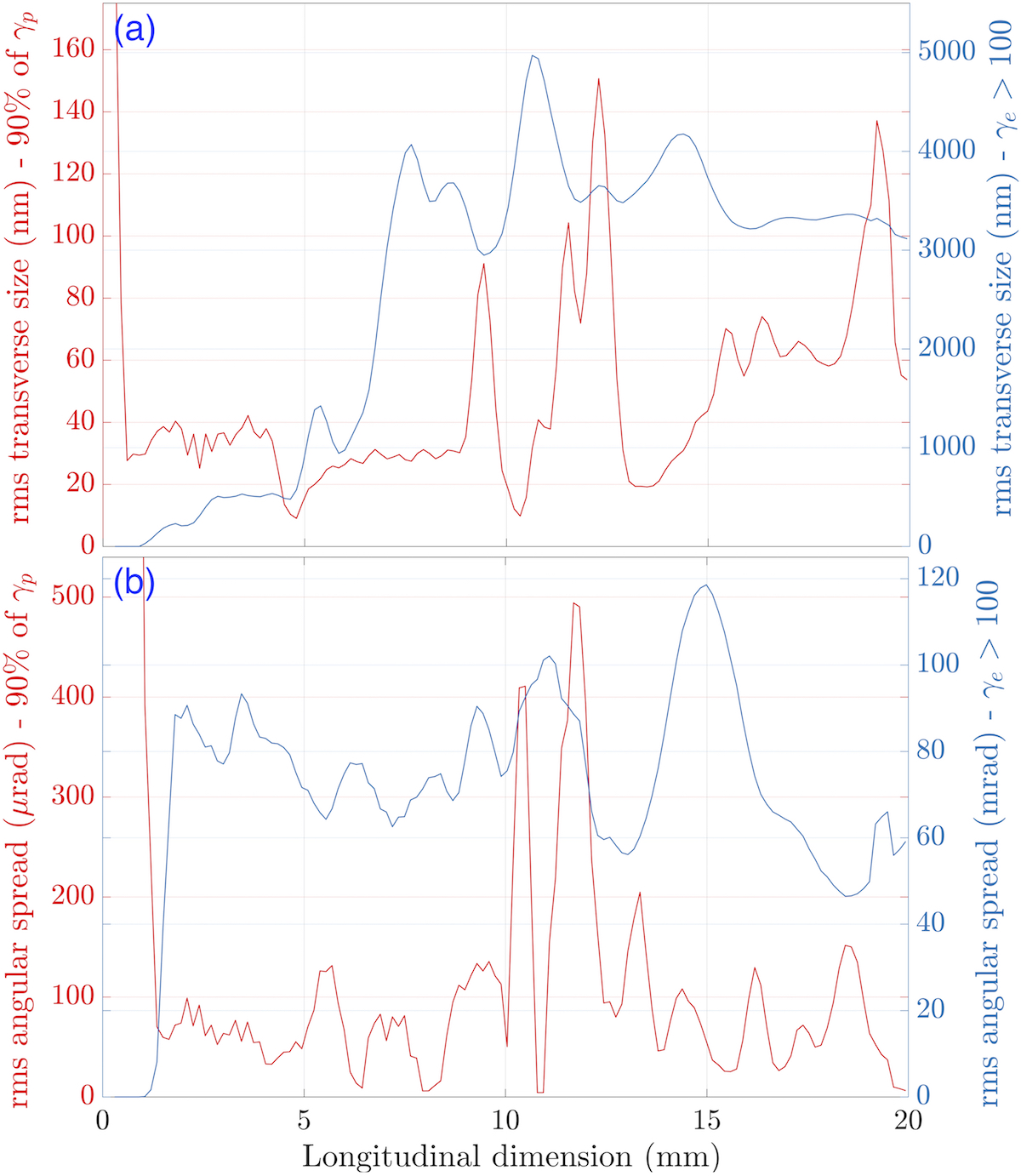}
   \caption{Beam root-mean-square (rms) transverse-size (in a) and angular-size (in b) evolution with laser propagation distance at $n_0 = 2\times 10^{18}\rm{cm^{-3}}$ from 2D PIC simulations.}
   \label{Fig8:pic-beam-property-evolution}
\end{figure}

In Fig.\ref{Fig7:peak-beam-energy-evolution}, the evolution of the peak Lorentz factor of the accelerated electrons or indirectly the kinetic energy of the most energetic electrons is shown. It is observed that the peak Lorentz factor clearly has two distinct phases which are represented in the two {\it different slopes} of the parabolic evolution of peak Lorentz factor with the longitudinal dimension. The parabolic shape is the characteristic of the integration over a linearly decreasing accelerating electric field magnitude as the beam gains energy over the back half of the bubble ($\partial\gamma_p/\partial\xi \propto e \times E^0_{\parallel}(1-\xi/R_B(\xi,t))$, where $E^0_{\parallel}$ is the peak field in the back of bubble). 

The first energy-gain slope is observed until around 25ps. The energy gain of the first set of injected particles stops as they over-run the accelerating phase of the wake due to bubble expansion. The second slope is observed after 25ps (after optical shock driven injection event around 24ps) where the peak Lorentz factor rises to $\gamma_p > 3000$ for $n_0 = 1.5 ~ - ~ 3.0 \times 10^{18} \rm{cm^{-3}}$ and $\gamma_p > 4000$ for $n_0 \simeq 2 \times 10^{18} \rm{cm^{-3}}$. The second energy gain slope is much higher due to the larger peak plasma fields excited by the strong optical-shock. In the matched regime (shown with a black dotted line), the maximum energy gain is limited to less than 2GeV.

In Fig.\ref{Fig8:pic-beam-property-evolution}(a), the root-mean-square (rms) beam transverse size and Fig.\ref{Fig8:pic-beam-property-evolution}(b), the root-mean square beam angle that are plotted, correspond to optimum energy gain case for $n_0 = 2 \times 10^{18} \rm{cm^{-3}}$. There are two different metrics which are plotted for both these parameters. These two metrics are necessary because the accelerated electrons are spread over a large volume of the full 6D phase-space and it is hard to properly define $\sigma_r$ and $\sigma_{\theta}$. The first metric adopted accounts all the particle with $\gamma>100$ and second metric only looks at all the particles within $\gamma > 0.9\times\gamma_p$. So, the second metric characterizes the properties of the highest energy particles.

The rms-size in the y-direction for the component of the beam with the highest energy particles (within 90\% of $\gamma_p$, where $\gamma_p$ evolves as shown in Fig.\ref{Fig7:peak-beam-energy-evolution}) is around an average of 50 nm, through most of the evolution. However, around the exit the beam transverse size is closer to 100nm for the multi-GeV beam ($\gamma_p \simeq 5500$ for $\rm n_0=2\times 10^{18}\rm{cm^{-3}}$ at an interaction length of 20mm). The rms-size in the y-direction for all the particles with $\gamma>100$ is on average around 3.5 $\mu m$.

The rms-angle (using $\sqrt{\langle p_y^2 \rangle / \langle p_x^2 \rangle} \equiv \sqrt{ \langle p_{\perp}^2 \rangle / \langle p_{\parallel}^2 \rangle }$) for the component of the beam containing the highest energy particles is around 100 $\mu \rm{rad}$. Whereas the rms-angle for the particles with $\gamma>100$ is on average around 80 mrad.

The estimated effective geometrical emittance for the high-energy component of the beam is $\rm\varepsilon_p = \text{rms-y}_p \times \text{rms-}\theta_p \simeq 10^{-5}$ mm-mrad. This corresponds with normalized emittance $\varepsilon_{p-n} = \gamma_p \times \varepsilon_p = 0.04 \text{mm-mrad}$. 

These observation of distinct properties of the particles at the peak energy in comparison to all particle above 50 MeV points towards the {\it adiabatic damping} effect of the geometrical emittance of particles as they undergo acceleration in the plasma. It is also of interest to note the conservation of the emittance of the highest energy particles of the beam, which is seen in the anti-correlation between the $\text{rms-y}_p$ and $\text{rms-}\theta_p$ in Fig.\ref{Fig8:pic-beam-property-evolution}(a) and \ref{Fig8:pic-beam-property-evolution}(b). Thus, the multi-GeV component of the laser-plasma accelerated beam in the strongly mismatched behaves like a high-quality conventional particle beam.

\section{Conclusion}\label{sec:conclusion}
A strongly mismatched regime that underlies many ground-breaking self-guided LPA experiments is modeled for the first-time using theoretical and computational analysis. The physical mechanisms that underpin the laser evolution, beam injection and acceleration in this regime are shown to significantly differ from the well established perfectly matched regime model. 

Two new signature physical processes of this regime, optical shock excitation and bubble elongation, have been investigated. The excitation of a strong optical-shock is shown to lead to the second signature process of rapid bubble elongation due to the longitudinal ponderomotive force far exceeding the radial force. A novel self-injection method due to bubble elongation is shown to inject a high-quality beam on-axis with unique properties.

Therefore launching larger focal-spot laser pulses in the strongly mismatched self-guiding regime based upon the underlying acceleration mechanisms uncovered here is a novel approach to produce high-energy beams with large self-injected charge of high transverse quality and higher overall laser to beam efficiency. 

Electron beams of this type will be useful for future work on laser positron acceleration and other applications that do not require a quasi-monoenergetic electron beam.
\section*{Acknowledgement}
The EPOCH PIC code used here is acknowledged.


\begin{thebibliography}{99}

\bibitem{Tajima-Dawson}
Tajima, T., Dawson, J. M., 
	\href{http://link.aps.org/doi/10.1103/PhysRevLett.43.267}{Phys. Rev. Lett. {\bf 43}, pp.267-270 (1979)}, doi: 10.1103/PhysRevLett.43.267.
	
\bibitem{Pukhov-bubble-2002}
Pukhov, A., Meyer-ter-Vehn, J.,
	\href{http://dx.doi.org/10.1007/s003400200795}{J. Appl Phys B {\bf 74}, iss.4-5, pp.355-361 (2002).}, doi: 10.1007/s003400200795.
		
\bibitem{Mangles-2004}
Mangles, S. P. D., Murphy, C. D., Najmudin, Z., Thomas, A. G. R., Collier, J. L., Dangor, A. E., Divall, E. J., Foster, P. S., Gallacher, J. G., Hooker, C. J., et. al.,
	\href{http://dx.doi.org/10.1038/nature02939}{Nature {\bf 431}, 535 (2004)}, doi:10.1038/nature02939.
	
\bibitem{Faure-2004}
Faure, J., Glinec, Y., Pukhov, A., Kiselev, S., Gordienko, S., Lefebvre, E., Rousseau, J.-P., Burgy, F.  and Malka, V.,
	\href{http://dx.doi.org/10.1038/nature02963}{Nature {\bf 431}, 541 (2004)}, doi:10.1038/nature02963.
	
\bibitem{UT-Austin}
Wang, X., Zgadzaj, R., Fazel, N., Li, Z., et. al.,
	\href{https://doi.org/10.1038/ncomms2988}{Nature Communications {\bf 4}, 1988 (2013)}
	
\bibitem{Nebraska}
Golovin, G., Chen, S., Powers, N., Liu, C., Banerjee, S., et. al.
	\href{https://doi.org/10.1103/PhysRevSTAB.18.011301}{Phys. Rev. ST Accel. Beams {\bf 18}, 011301 (2015)}
	
\bibitem{GIST}
Kim, H. T., Pathak, V. B., Pae, K. H., Lifschitz, A., et. al.,
	\href{https://doi.org/10.1038/s41598-017-09267-1}{Scientific Reports {\bf 7}, 10203 (2017)}
	
\bibitem{Strathclyde}
E. Brunetti, R. P. Shanks, G. G. Manahan, M. R. Islam, B. Ersfeld, et. al.,
	\href{https://doi.org/10.1103/PhysRevLett.105.215007}{Phys. Rev. Lett. {\bf 105}, 215007 (2010)}
	
\bibitem{Lu_PRSTAB_2007}
Lu, W., Tzoufras, M., Joshi, C., Tsung, F. S., Mori, W. B., Vieira, J., Fonseca, R. A., Silva, L. O., 
	\href{http://dx.doi.org/10.1103/PhysRevSTAB.10.061301}{Physical Review Special Topics - Accelerators and Beams 10, 061301 (2007)}, doi:10.1103/PhysRevSTAB.10.061301
	
\bibitem{positron-LPA}
Sahai, A. A., 
	\href{https://doi.org/10.1103/PhysRevAccelBeams.21.081301}{Phys. Rev. Accel. Beams 21, 081301 (2018)}

\bibitem{Siegman-1997}
Siegman, A. E.,
	\href{http://dx.doi.org/10.1117/12.150601}{Proc. SPIE 1868, Laser Resonators and Coherent Optics: Modeling, Technology, and Applications, 2 (1993)}, doi 10.1117/12.150601;
Siegman, A. E.,
	\href{https://doi.org/10.1364/DLAI.1998.MQ1}{DPSS (Diode Pumped Solid State) Lasers: Applications and Issues, Optical Society of America, Vol. 17,1998, paper MQ1}
	
\bibitem{Strehl-OL-1998}
J-C. Chanteloup, F. Druon, M. Nantel, A. Maksimchuk, G. Mourou,
	\href{https://doi.org/10.1364/OL.23.000621}{Optics Letters {\bf 23}, iss. 8, pp. 621-623 (1998)};
Druon, F., Cheriaux, G., Faure, J., Nees, J., Nantel, M., Maksimchuk, A., Mourou, G., Chanteloup, J-C., Vdovin, G.,
	\href{https://doi.org/10.1364/OL.23.001043}{Opt. Lett. {\bf 23}, pp.1043-1045 (1998)}

\bibitem{Max-1974}
Max, C. E., Arons, J., Langdon, A. B.,
	\href{https://doi.org/10.1103/PhysRevLett.33.209}{Phys. Rev. Lett. {\bf 33}, 209, (1974)}, doi: 10.1103/PhysRevLett.33.209.
	
\bibitem{Sprangle-1987}
Sprangle, P., Cha-Mei Tang ; E. Esarey
	\href{http://dx.doi.org/10.1109/TPS.1987.4316677}{IEEE Transactions on Plasma Science, {\bf 15}, iss. 2, pp.145-153 (1987)}, doi: 10.1109/TPS.1987.4316677.
	
\bibitem{Sun-1987}
Sun, G. Z., Ott, E., Lee, Y. C., Guzdar, P., 
	\href{http://dx.doi.org/10.1063/1.866349}{Phys. of Fluids {\bf 30}, 526 (1987)}, doi: 10.1063/1.866349.
	
\bibitem{Hafizi-2000}
Hafizi, B., Ting, A., Sprangle, P., Hubbard, R. F.,
	\href{https://journals.aps.org/pre/pdf/10.1103/PhysRevE.62.4120}{Phys. Rev. E 62, 4120, (2000)}, doi:10.1103/PhysRevE.62.4120.

\bibitem{KV-1959}	
Kapchinsky, I. M. and Vladimirsky, V. V., 
	\href{http://inspirehep.net/record/919865/files/HEACC59_297-311.pdf}{Proc. Int. Conf. on High-Energy Accelerators and Instrumentation, CERN, Geneva, pp. 274-288, (1959)}.
	
\bibitem{Kalmykov-2009}
Kalmykov, S., Yi, S. A., Khudik, V., Shvets, G.,
	\href{https://doi.org/10.1103/PhysRevLett.103.135004}{Phys. Rev. Lett. {\bf 103}, 135004 (2009)}
	
\bibitem{Poder-2016}
Poder, K.,
	\href{}{Ph.D. Thesis, Imperial College London, UK}
	
\bibitem{Kneip-2009}
Kneip, S., et. al.,
	\href{https://doi.org/10.1103/PhysRevLett.103.035002}{Phys. Rev. Lett. 103, 035002 (2009)}, doi:10.1103/PhysRevLett.103.035002
	
\bibitem{Faure-2015}
Beaurepaire, B., Vernier, A., Bocoum, M. , Bohle, F., Jullien, A., Rousseau, J.P., Lefrou, T. , Douillet, D., Iaquaniello, G., Lopez-Martens, R., Lifschitz, A., Faure, J.,
	\href{https://doi.org/10.1103/PhysRevX.5.031012}{Phys. Rev. X 5, 031012 (2015)}
	
\bibitem{Jacobi-elliptic-functions}
Jacobi, C. G. J., 
	 \href{https://archive.org/details/fundamentanovat00jacogoog}{Konigsberg, Germany: Regiomonti, Sumtibus fratrum Borntraeger, 1829};
Chandrasekharan, K., 
	Berlin: Springer-Verlag, 1985, ISBN-10: 0387152954
	
\bibitem{KdV-eq}
Korteweg, D. J. and de Vries, G., 
	\href{http://dx.doi.org/10.1080/14786449508620739}{Philosophical Magazine {\bf 39}, 240, pp.422-443, (1895).} doi:10.1080/14786449508620739
	
\bibitem{optical-shock}
Gaeta, A. L.,
	\href{https://doi.org/10.1103/PhysRevLett.84.3582}{Phys. Rev. Lett. 84, 3582, (2000)}, doi:10.1103/PhysRevLett.84.3582

\bibitem{etching-model-1996}
Decker, C. D., Mori, W. B., Tzeng, K.C., T. Katsouleas,
	\href{https://doi.org/10.1063/1.872001}{Physics of Plasmas 3, 2047 (1996)}
	
\bibitem{Ehrlich-1996}
Ehrlich, Y., Cohen, C., Zigler, A., Krall, J., Sprangle, P. and Esarey, E.,
	\href{https://doi.org/10.1103/PhysRevLett.77.4186}{Phys. Rev. Lett. 77, 4186, (1996)}
	
\bibitem{chen-lens-1986}
Chen, P.,
	\href{http://www.slac.stanford.edu/cgi-wrap/getdoc/slac-pub-4049.pdf}{SLAC-PUB-4049, August 1986}

\bibitem{EPOCH-paper}	
T D Arber, K Bennett, C S Brady, et. al. 
	\href{https://doi.org/10.1088/0741-3335/57/11/113001}{Plasma Phys. Control. Fusion {\bf 57}, 113001, (2015)}
	
\bibitem{sahai-thesis-2015}
Sahai, A. A., 
	\href{http://search.proquest.com/docview/1753637333}{Ph.D. dissertation, Duke university, July 2015}

\end{thebibliography}
\end{document}